\documentclass[12pt,preprint]{aastex}
\usepackage{amsmath, amsthm, amssymb}

\def\au{\,\textsc{au}}

\def\bfo{{\bf o}}

\def\half{\textstyle{\frac{1}{2}}}

\def\bfn{{\bf n}}

\def\bfz{{\bf z}}

\def\bfkappa{\mbox{\boldmath $\kappa$}}

\def\comment#1{}

\tighten

\begin{document}

\title{The statistics of multi-planet systems}

\author{Scott Tremaine and Subo Dong\altaffilmark{1} \\ Institute for Advanced Study,
  Princeton, NJ 08540, USA}

\altaffiltext{1}{Sagan Fellow}

\begin{abstract}

\noindent
We describe statistical methods for measuring the exoplanet
multiplicity function---the fraction of host stars containing a given
number of planets---from transit and radial-velocity surveys. The
analysis is based on the approximation of separability---that the
distribution of planetary parameters in an $n$-planet system is the
product of identical 1-planet distributions. We review the evidence
that separability is a valid approximation for exoplanets. We show how
to relate the observable multiplicity function in surveys with similar host-star
populations but different sensitivities. We also show how to correct
for geometrical selection effects to derive the multiplicity
function from transit surveys if the distribution of relative
inclinations is known. Applying these tools to the Kepler transit
survey and to radial-velocity surveys, we find that (i) the Kepler
data alone do not constrain the mean inclination of multi-planet
systems; even spherical distributions are allowed by the data but only
if a small fraction of host stars contain large planet populations
($\gtrsim 30$); (ii) comparing the Kepler and radial-velocity surveys
shows that the mean inclination of multi-planet systems lies in the
range 0--5 degrees; (iii) the multiplicity function of the Kepler
planets is not well-determined by the present data.

\end{abstract}

\section{Introduction}
\noindent
The distribution of inclinations in multi-planet systems provides
fundamental insights into planet formation. The small
inclinations of the planets in the solar system---the largest is
$7^\circ$, for Mercury---strongly suggest that they formed from a
disk. However, we should not be surprised if extrasolar planetary
systems have larger inclinations, for several reasons: (i) the rms
inclinations in the asteroid and Kuiper belts are substantially
larger, $12^\circ$ and $16^\circ$ respectively\comment{asteroids from
  Murison's database; all diameters > 20 km; in my file asteroids.dat;
  Kuiper belt from Brown AJ 121 2804, eq. 6}; (ii) in most
astrophysical disks, the rms eccentricity and inclination are
correlated, and the eccentricities of extrasolar planets are much
larger than those of solar-system planets (0.23 for exoplanets with
periods greater than 10 days, compared to 0.05); (iii) a number of
dynamical mechanisms can excite inclinations, including Kozai--Lidov
oscillations, planet-planet scattering, and resonance sweeping; (iv)
measurements of the Rossiter--McLaughlin effect in transiting systems
\citep[e.g.,][]{winn2010} show a broad distribution of obliquities
(angle between the spin axis of the host star and the orbit axis of
the planet) and some processes that excite obliquities do so by
exciting inclinations; (v) most extrasolar planetary systems have
quite different configurations from the solar system, so they may form
by quite different mechanisms; (vi) there are still serious
theoretical obstacles to the formation of planets from a circumstellar
disk, and several authors have suggested that some or all planets may
be formed by other mechanisms, more similar to star formation, that
would impart large inclinations to the planets 
\citep[e.g.,][]{b97,pt01,rme07,abt10}.

There is only fragmentary evidence that extrasolar
planetary systems have small relative inclinations:

\begin{itemize}

  \item The mutual inclination of planets B and C in the system
  surrounding the pulsar B1257+12 is less than $\sim 13^\circ$
  \citep{kw03}; this result is only marginally relevant to  planetary
  systems around main-sequence stars since pulsar planets must have
  had a very different history. 

  \item Using radial-velocity and astrometric data, \cite{bs09}
  estimate that the mutual inclination between GJ 876 b and c is
  $5.0^\circ{+ 3.9^\circ\atop -2.3^\circ}$. Using radial
  velocities and dynamical modeling of the planet-planet interactions
  \cite{cor10} conclude that the mutual inclination is $\lesssim
  2^\circ$, while \cite{bal11} finds that the same quantity is between
  5 and $15^\circ$. The large scatter among these results means that
  they should be used with caution. 

  \item \cite{mc10} find from astrometric and radial-velocity
  measurements that the mutual inclination of $\upsilon$ And c and d
  is $30^\circ\pm1^\circ$, much larger than in GJ 876 but still small
  enough to suggest formation from a disk. 

  \item Dynamical fits to the transit timing of two planets in the
  Kepler-9 system yield an upper limit to the mutual inclination of
  $\sim10^\circ$ \citep{hol10}. However, this system was discovered in
  a transit survey, and such surveys are far more likely to detect
  multi-planet systems with small inclinations rather than large ones.

\item \cite{liss11a} studied the six-planet system Kepler-11 and
  concluded that the absence of transit duration changes in Kepler-11e
  implies that its inclination relative to the mean orbital plane of
  other planets is less than 2 degrees\footnote{Lissauer et al.\ also
    concluded that the mean mutual inclination of the planets was
    1--$2^\circ$ from Monte Carlo simulations of the probability that
    a randomly placed observer would see transits of all the planets;
    however, this conclusion is suspect since the probability that a
    random star with six planets would show six transits is different
    from the probability that one star from the Kepler sample of
    $\sim150,000$ stars would show six transits.}; once again, this
  result is biased by the strong dependence of the probability that
  two or more planets will transit on their mutual inclination.

\end{itemize}

As noted above, if one planet in a two-planet system transits its host star as viewed
from Earth, the probability that the second planet will also transit
is higher if the mutual inclination of the two planetary
orbits is small \citep[e.g.,][]{rh10}. This argument suggests that the 
numbers of 1-planet, 2-planet,$\ldots$,$N$-planet systems detected in a
large transit survey contain information about both the multiplicity
function---the fraction of host stars containing
$0,1,2,\ldots,N$ planets---and the inclination distribution. The
challenge is to disentangle the two distributions to
distinguish thick systems with many planets from thin systems with few
planets. 

The first attempt to do this was made by \cite{liss11b}, who modeled
the number of multiple-planet systems detected in the first four
months of data from the Kepler survey \citep{bor11}---115 with two
transiting planets, 45 with three, 8 with four, and one each with 5
and 6. Lissauer et al.\ used a variety of simple models for the
distribution of the number of planets per system. They found that none of their
models fit the data well, mostly because they produced too few systems
in which a single transiting planet was observed, but that the
best-fit models typically had mutual inclinations $\lesssim 5^\circ$.

The purpose of this paper is to develop a general formalism that
relates the intrinsic properties of multi-planet systems to the
properties of the multi-planet systems that are detected in transit or
other surveys (\S\ref{sec:survey} and \S\ref{sec:transit}), and to
apply this formalism to the Kepler planet survey (\S\ref{sec:kepler})
and to radial-velocity surveys (\S\ref{sec:keprv}). Previous
analyses have used Monte Carlo simulations to explore these problems,
but our calculations are mostly analytic or semi-analytic and do not employ Monte Carlo
methods.

\subsection{Preliminaries}

\noindent
First we introduce some notation. (i) The Kepler team uses the term
planet ``candidate'' to denote a possible planet that has
been discovered through transits but not yet been confirmed
by radial-velocity measurements. \cite{mor11} estimate that
90\% to 95\% of the Kepler planet candidates are real planets, so for
the remainder of this paper we will simply assume that all the Kepler
planet candidates are real and delete the word
``candidate''. (ii) We must constantly distinguish between the number
of planets in a system and the number of transiting planets in that
system. We use the contraction ``tranet'' to denote ``transiting
planet''. Thus one could have, for example, a two-tranet, three-planet
system (\citealt{rh10} call this a ``double-transiting triple system'').
(iii) We distinguish two types of selection effects that limit a planet
sample.  Every survey has a set of detection thresholds, determined by the
parameters of the survey, that limit the properties of the planets
that it can detect (maximum orbit period, minimum
reflex radial velocity, minimum transit depth, etc.). A {\em survey}
selection effect is a limitation on the number of detectable planets
due to the detection thresholds. A {\em geometrical} selection effect
is a limitation arising from the orientation of the planetary
system---in particular, the planet must cross in front of the stellar
disk to be detectable in a transit survey\footnote{There is also a
  geometrical selection effect in radial-velocity surveys, since the
  reflex velocity is proportional to $\sin \gamma$ where $\gamma$ is
  the inclination of the planetary orbit to the line of
  sight. However, we can eliminate this effect by working only with
  the minimum mass $M\sin \gamma$ where $M$ is the planet
  mass; of course, for transit surveys $\sin \gamma\simeq 1$ so the
  minimum mass equals the mass.}.

We assume that the stars in a survey may have $0,1,\ldots,K$ planets
and denote the number of stars in the survey with $k$ planets by
$N_k$.  Thus $\sum_{k=0}^KN_k$ is the total number of stars in the
survey. The vector $\mathbf{N}=(N_0,N_1,\ldots,N_K)$ is called the
multiplicity function. 

Because of survey and geometric selection effects, only a fraction of
these planets will be detected in the survey. Let the survey selection
matrix element $S_{km}$ be the
probability that a system containing $m$ planets has $k$ of them that
pass the survey selection criteria. Similarly, let the geometric
selection matrix $G_{jk}$ be the
probability that $j$ of these $k$ planets pass the geometric selection
criteria. Then the expected number of systems that the
survey should detect with $j$ tranets is 
\begin{equation}
\overline{n}_j=\sum_{k=j}^K
G_{jk}\sum_{m=k}^KS_{km}N_m,\quad \mbox{or}\quad \overline{\mathbf{n}}=\mathbf{G}\cdot\mathbf{S}\cdot\mathbf{N}.
\label{eq:gdefqq}
\end{equation}
We call $\overline{\mathbf{n}}$ the observable multiplicity function. 
Clearly
\begin{equation}
G_{mn}=S_{mn}=0\hbox{
  for }m>n,\quad G_{00}=S_{00}=1, \quad G_{mn},S_{mn}\ge 0.
\label{eq:G}
\end{equation}
Moreover since the number of detectable planets in an $n$-planet
system must be between 0 and $n$, we have 
\begin{equation}
\sum_{m=0}^nG_{mn}=\sum_{m=0}^nS_{mn}=1.
\label{eq:sumG}
\end{equation}
Thus $\mathbf{G}$ and $\mathbf{S}$ are $(K+1)\times(K+1)$
upper-triangular stochastic matrices. For physical reasons
$\mathbf{G}$ and $\mathbf{S}$ should commute (eq.\ \ref{eq:gdefqq}
should not depend on whether we consider the survey selection effects
or the geometric selection effects first). We have confirmed that 
the commutator $[\mathbf{G},\mathbf{S}]$ is indeed zero for the
selection matrices that we derive below.

\subsection{Separability}

\label{sec:separable}

\noindent
Let $\mathbf{w}$ represent all of the orientation-independent
properties of a planet and its host star that determine its
detectability (planetary mass and radius; stellar mass, radius,
distance, and luminosity; orbital period, etc.) and let
$f(\mathbf{w}_1,\ldots,\mathbf{w}_n)$ represent the probability distribution of these
parameters for an $n$-planet system. Thus $\int d\mathbf{w}_1\cdots
d\mathbf{w}_n\,f(\mathbf{w}_1,\ldots,\mathbf{w}_n)=1$. 

A natural assumption for describing multi-planet systems
is that the $n$-planet distribution function is {\em separable}, that
is,
\begin{equation}
f(\mathbf{w}_1,\ldots,\mathbf{w}_n)=\prod_{m=1}^n f(\mathbf{w}_m),
\qquad \int d\mathbf{w}\,f(\mathbf{w})=1.
\label{eq:sep}
\end{equation}
This assumption can only be approximately valid---for example, it is
inconsistent with the observational finding that planets tend to be
concentrated near mutual orbital resonances, and with the theoretical
finding that planets separated by less than a few Hill radii are
unstable. Nevertheless, we argue that the separability assumption
is sufficiently accurate to provide a powerful tool for analyzing the
statistics of multi-planet systems. We describe the evidence on its
validity in \S\ref{sec:valid}.

\section{Survey selection effects}

\label{sec:survey}
Let $\Theta^{A}(\mathbf{w})$ be the probability that a planet
with properties $\mathbf{w}$ is detected in the survey labeled by A if
its host star is on the target list for this survey and the
orientation of the observer is correct (we assume that whether or not
a planet can be detected is independent of the presence or absence of
other planets in the same system, which is a reasonable first
approximation). Thus the function $\Theta^{A}(\mathbf{w})$
describes the survey selection effects for A, but not the geometric
selection effects. The probability that a planet is detected, ignoring
geometric selection effects, is then
\begin{equation}
W^{A}=\int
f(\mathbf{w})\Theta^{A}(\mathbf{w})\,d\mathbf{w}.
\label{eq:wdef}
\end{equation}
If the survey target list contains $N^{A}_m$ stars with $m$ planets, then
using the separability assumption (\ref{eq:sep}) the expected number of systems
in which $k$ planets will be detected is
\begin{equation}
  \overline{n}^{A}_k=\sum_{m=k}^K
  S_{km}(W^{A})N^{A}_m, \quad
  0\le k\le K;
\label{eq:ccc}
\end{equation}
where the survey selection matrix $\mathbf{S}$ is a $(K+1)\times(K+1)$
matrix whose entries are given by the binomial distribution,
\begin{equation}
S_{km}(W)\equiv \frac{m!}{k!(m-k)!}W^k(1-W)^{m-k}, \quad 0\le k\le
m\le K,
\label{eq:sdef}
\end{equation}
and zero otherwise.  Note that $\mathbf{S}(1)$ is the unit matrix. A
useful identity is \citep[e.g.,]{strum}
\begin{equation}
\mathbf{S}(A)\cdot\mathbf{S}(B)=\mathbf{S}(AB),
\label{eq:mult}
\end{equation}
which in turn implies
\begin{equation}
\mathbf{S}^{-1}(W)=\mathbf{S}(W^{-1}).
\label{eq:inverse}
\end{equation}
Although the physical motivation (\ref{eq:ccc}) for the definition of
$\mathbf{S}(W)$ requires $0\le W\le 1$, the matrix is well-defined for
all values of $W$.

With the assumption of separability it is straightforward to show that
the conditional probability distribution of the parameters
$\mathbf{w}_m$, given that $k$ planets are detected, is (cf.\ eq.\
\ref{eq:sep})
\begin{equation}
f(\mathbf{w}_1,\ldots,\mathbf{w}_k)=\prod_{m=1}^k
f(\mathbf{w}_m).
\label{eq:seprv}
\end{equation}
Thus a separable distribution is still separable after survey
selection effects are applied, so long as the selection effects depend
only on the properties of an individual planet. 

The factor $W$ (eq.\ \ref{eq:wdef}) is usually difficult to determine
reliably since (i) we do not have good models for the distribution
$f(\mathbf{w})$ of the planetary parameters; (ii) in most cases the
survey selection effects $\Theta(\mathbf{w})$ are not known
accurately; (iii) in many cases the target list from which a given
sample of exoplanets was detected is not even known (the Kepler survey
is an exception to the last two limitations). However, useful results
can be obtained without an explicit evaluation of $W$. Suppose, for
example, we have two surveys A and B that examine populations of
target stars with similar characteristics; then the ratio of the
number of $m$-planet systems in the target populations of the two
surveys should be independent of $m$, so
$N^{B}_m=cN^{A}_m$ where $c$ is a constant given by the
ratio of the number of target stars in B and A. Equation
(\ref{eq:ccc}) can then be written
\begin{equation}
  \overline{\mathbf{n}}^{A}=\mathbf{S}(W^{A})\mathbf{N}^{A}, \quad
  \overline{\mathbf{n}}^{B}=c\,\mathbf{S}(W^{B})\mathbf{N}^{A}.
\label{eq:cccc}
\end{equation}
Applying equations (\ref{eq:mult}) and (\ref{eq:inverse}), we have
\begin{equation}
  \overline{\mathbf{n}}^{B}=c\,\mathbf{S}(f^{BA})\overline{\mathbf{n}}^{A}
\label{eq:ccccc}
\end{equation}
where $f^{BA}\equiv W^{B}/W^{A}=1/f^{AB}$.
Thus the observable multiplicity function $\overline{\mathbf{n}}^{B}$ of
survey B is directly related to that of survey A by a matrix that
depends only on a single parameter $f^{BA}$ (the normalization
constant $c$ is known, since it is just the ratio of the number of
target stars in the two surveys). The parameter $f^{BA}$ can be
eliminated if we plot
$\overline{n}_2^{B},\overline{n}_3^{B},\ldots$ as
functions of $\overline{n}_1^{B}$. In practice we must use the
multiplicity function $\mathbf{n}^{A}$ rather than $\overline{\mathbf{n}}^{A}$ on the
right side of equation (\ref{eq:ccccc}) but these should not be very
different so long as $n_k^{A}\gg1$. Equation (\ref{eq:cccc}) is
well-defined whether $f^{BA}$ is larger or smaller than unity,
but if $f^{BA}>1$ the statistical errors will be amplified and
it is likely that some of the predicted values of
$\overline{n}_k^{B}$ will be negative, which is
unphysical. Thus, if the separability approximation is valid, the
observable multiplicity function of deep surveys can be used to predict the
observable multiplicity function of shallow surveys (but not vice versa).

\begin{figure}
\centering
\includegraphics[clip=true,width=0.95\hsize]{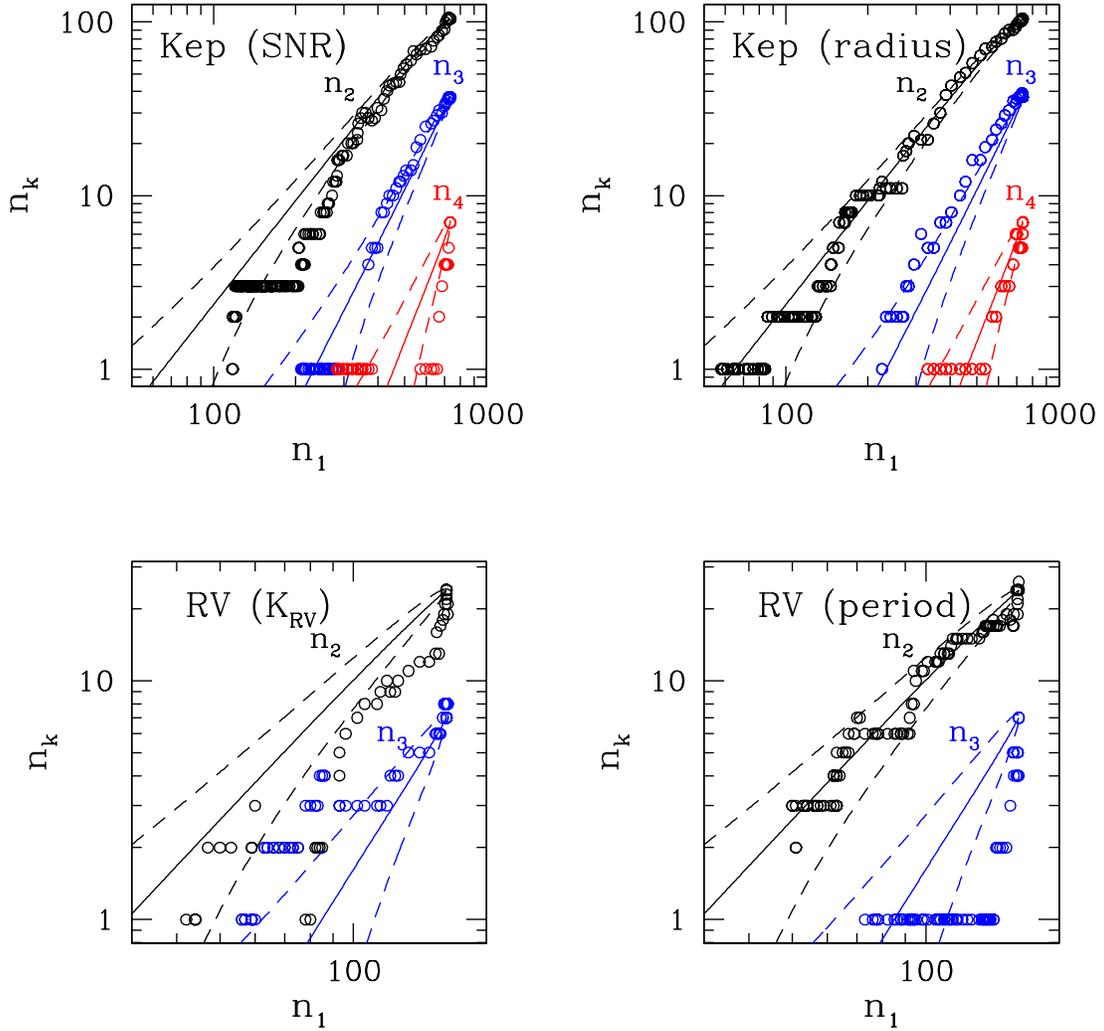}
\caption{The observable multiplicity function for subsets of the Kepler and
  radial-velocity planet samples (top and bottom panels,
  respectively). The catalog subsets are defined by imposing cuts
  based on signal-to-noise ratio (SNR), planet radius, orbital period, or
  velocity semi-amplitude ($K_\mathrm{RV}$). Open circles show $n_2$, $n_3$, $n_4$
  (numbers of 2, 3, and 4-planet systems) as a function of
  $n_1$. Solid and dashed curves show the predictions of equation
  (\ref{eq:ccccc}) and the 1--$\sigma$ errors on the predictions.}\label{fig:one}
\end{figure}

\paragraph*{Example} To illustrate this procedure, we examine the
Kepler catalog of \cite{bor11}, trimmed by 20\% as described at
the start of \S\ref{sec:kepler} to produce a more homogeneous
set of target stars. This is catalog A. All of the planets in this catalog are detected
with a signal/noise ratio (SNR) of at least 7. We construct a sequence
of shallower catalogs (catalogs ``B'') by gradually increasing the
minimum SNR up to values exceeding 100, at which point only a handful
of multi-planet systems is left. The relation (\ref{eq:ccccc}) implies
that apart from statistical fluctuations the numbers of
multiple-planet systems $n_k^{B}$, $k=1,\ldots$, are functions
only of $f^{BA}$ and the known $\mathbf{n}^{A}$, which
approximates the observable multiplicity function
$\overline{\mathbf{n}}^{A}$. Hence by eliminating $f^{BA}$ in favor
of $n_1^{B}$, the number of $k$-planet systems in any survey B
can be predicted as a function of the number of one-planet systems in
that survey. These predictions for $k=2,3,4$ are shown in the upper
left panel of Figure \ref{fig:one} as solid lines, along with the
1--$\sigma$ confidence bands (dashed lines). The actual numbers of
multi-planet systems after SNR cuts on the Kepler data are shown as
open circles. Within the statistical errors the predictions agree with
the data for $k=2$ and 3 and are marginally consistent for $k=4$:
using a Kolmogorov--Smirnov (KS) test\footnote{The use of a KS test is
  not strictly applicable since $n_k$ and $n_1$ are cumulative
  distributions of a third parameter, the SNR, rather than being
  directly related. However, the results should be approximately
  correct when $n_1\gg n_k$ which is usually the case.}, the
$p$-value (probability of observing deviations at least as
extreme as those seen, given the null hypothesis) is 0.28, 0.27, and
0.06 respectively. 

The upper right panel of Figure \ref{fig:one} shows a similar
comparison for a sequence of catalogs based on cuts at increasing planet radius,
rather than SNR. The results are consistent with the separable model
to within the statistical errors ($p$-values of 0.72, 0.19,
and 0.66 for $k=2,3,4$). 

The lower panels of Figure \ref{fig:one} show similar results for radial-velocity
surveys. The ``A'' catalog consists of 240 FGK dwarf stars
hosting one or more planets (see eq.\ \ref{eq:plnumrv} for more detail), and the cuts are based on $K_\mathrm{RV}$
(semi-amplitude of the radial-velocity curve) on the left and orbital
period on the right. The predictions are marginally
consistent with the null hypothesis ($p$-values between 0.03
and 0.10) except for $n_2$ as a function of the cut in $K_\mathrm{RV}$, for which
the null hypothesis is excluded. 

These results confirm that in many cases the separability
approximation and equation (\ref{eq:ccccc}) provide useful tools for
removing survey selection effects and converting the observable multiplicity
function between surveys.

\section{Geometric selection effects in transiting systems}

\label{sec:transit}

\noindent
Throughout this paper we shall assume that tranets are in circular
orbits. \cite{moor11} estimate that the mean eccentricity of planets
discovered in the Kepler survey is only 0.1--0.25, so this assumption
should not cause significant errors. We shall also
assume that a transit occurs when the line of sight to the center of
the planet intersects the stellar disk. This assumption should be
approximately correct so long as the planetary radius is much smaller
than the stellar radius (the median ratio of planetary radius to
stellar radius in the Kepler survey is only 0.026).

Let $R_\star$ be the radius of the star, $a$ the semi-major axis of a
planet in a circular orbit, and $\epsilon\equiv R_\star/a$. Consider a
system containing $n$ planets with semi-major axes specified by
$\epsilon_1,\ldots,\epsilon_n$.  Let
$g_{mn}(\epsilon_1,\ldots,\epsilon_n)$ be the probability that a
randomly oriented observer will detect $m$ tranets in this system.

\paragraph{One planet} First consider the case $n=1$. We define three
unit vectors: $\hat\bfo$ points towards the observer, $\hat\bfn$ is
normal to the planetary orbit, and $\hat\bfz$ is normal to the
reference plane from which inclinations $i$ are measured. Thus
$\hat\bfz\cdot\hat\bfn=\cos i$ and $\hat\bfo\cdot\hat\bfn=\cos
\gamma$. If the planet's size is negligible, it transits if and only
if $|\hat\bfo\cdot\hat\bfn|<\epsilon$ or
$|\cos\gamma|<\epsilon$ so
\begin{equation} 
g_{11}(\epsilon)=1-g_{01}(\epsilon)=\frac{\int_{|\cos\gamma|<\epsilon}\sin\gamma\,
d\gamma}{\int\sin\gamma \,d\gamma}=\epsilon.
\label{eq:goneone}
\end{equation}

\paragraph{Two planets} Let $h(w)=1$ if $|w|<1$ and zero otherwise.
Then transits occur if and only if $h(\epsilon^{-1}\cos\gamma)$ is unity and
we may write
\begin{equation} 
h(\epsilon^{-1}\cos\gamma)=\sum_{\ell=0}^\infty b_\ell(\epsilon) P_\ell(\cos\gamma)
\end{equation}
where $P_\ell$ is a Legendre polynomial. From the properties of these functions we have
\begin{equation}
b_\ell(\epsilon)=\left\{\begin{array}{ll}\epsilon, & \ell=0, \\
P_{\ell+1}(\epsilon)-P_{\ell-1}(\epsilon), &
  \ell\hbox{ even, } \ell>0 \\
              0, & \ell\hbox{ odd.}\end{array}\right.
\end{equation}
Now let $(\theta,\phi)$ be the polar coordinates for $\hat\bfo$ relative to
the polar axis $\hat\bfz$, and $(\Omega-\half\pi,i)$ the polar coordinates for
$\hat\bfn$. Then
\begin{equation} 
h(\epsilon^{-1}\cos\gamma)=4\pi\sum_{\ell=0}^\infty \frac{b_\ell(\epsilon)}{2\ell+1}\sum_{m=-\ell}^\ell Y_{\ell m}^\ast(\theta,\phi)Y_{\ell
  m}(i,\Omega-\half\pi). 
\end{equation}
Let the probability distribution of planetary inclinations be
$q(i|\bfkappa)di$, where $\bfkappa$ is a
set of free parameters describing the inclination distribution, which
we may vary to fit the observations. Then the probability of a transit
of a single planet, given the observer orientation $x\equiv
\cos\theta$, is
\begin{align}
u(x|\epsilon,\bfkappa)=\int \frac{di\, d\Omega}{2\pi}q(i|\bfkappa)h(\epsilon^{-1}\cos\gamma) =&
4\pi\sum_{\ell=0}^\infty \frac{b_\ell(\epsilon)}{2\ell+1}\int di\,q(i|\bfkappa)Y_{\ell 0}^\ast(\theta,0)Y_{\ell
  0}(i,0)\notag \\
=& \sum_{\ell=0}^\infty Q_\ell(\bfkappa)\,b_\ell(\epsilon) P_\ell(x).
\label{eq:one}
\end{align}
where 
\begin{equation}
Q_\ell(\bfkappa)\equiv \int_0^\pi di\,q(i|\bfkappa)P_\ell(\cos i), \quad Q_0=1.
\label{eq:onetwo}
\end{equation}

If a system contains two planets, the probability that both transit
for a random orientation of the observer is 
\begin{align}
g_{22}(\epsilon_1,\epsilon_2,\bfkappa)=&\half \int_{-1}^1 dx\, u(x|\epsilon_1,\bfkappa)u(x|\epsilon_2,\bfkappa) \notag \\
=& \half\sum_{\ell,n=0}^\infty
b_\ell(\epsilon_1)b_n(\epsilon_2)Q_\ell(\bfkappa) Q_n(\bfkappa)\int_{-1}^1 dx\,P_\ell(x)P_n(x) \notag \\ 
=&\sum_{\ell=0}^\infty \frac{Q_\ell^2(\bfkappa)}{2\ell+1}b_\ell(\epsilon_1)b_\ell(\epsilon_2).
\end{align}
Moreover the probability that one and only one of the two planets transits is
\begin{align}
g_{12}(\epsilon_1,\epsilon_2,\kappa)=& \half \int_{-1}^1 dx\,\left\{u(x|\epsilon_1,\bfkappa)[1-u(x|\epsilon_2,\bfkappa)]
  +[1-u(x|\epsilon_1,\bfkappa)]u(x|\epsilon_2,\bfkappa)]\right\}\notag \\ 
=&g_{11}(\epsilon_1,\bfkappa)+g_{11}(\epsilon_2,\bfkappa)-2g_{22}(\epsilon_1,\epsilon_2,\bfkappa)
\label{eq:g12}
\end{align}
and the probability that no planets transit is
\begin{align}
g_{02}(\epsilon_1,\epsilon_2,\bfkappa)=&1-g_{12}(\epsilon_1,\epsilon_2,\bfkappa)
-g_{22}(\epsilon_1,\epsilon_2,\bfkappa)  \notag \\ =& 
1-g_{11}(\epsilon_1,\bfkappa)-g_{11}(\epsilon_2,\bfkappa)+g_{22}(\epsilon_1,\epsilon_2,\bfkappa).
\label{eq:g02}
\end{align}

For example, if the planets are distributed isotropically then
$q(i)di=\half\sin i\,di$, $Q_\ell=\delta_{\ell0}$ and
$g_{22}(\epsilon_1,\epsilon_2)=\epsilon_1\epsilon_2$. If the planets have zero
inclination, it can be shown that 
\begin{equation}
g_{22}(\epsilon_1,\epsilon_2)=\sum_{\ell=0}^\infty
\frac{b_\ell(\epsilon_1)b_\ell(\epsilon_2)}{2\ell+1}=\hbox{min\,}(\epsilon_1,\epsilon_2),
\label{eq:gtwotwo}
\end{equation}
although this result is derived more easily in other ways. 

\paragraph{Three or more planets} These results can be extended to any
number of planets\footnote{For $n=3$ the functions $g_{mn}$ can be
  expressed as series in the Wigner 3-$j$ symbols, but in practice it
  is simpler to evaluate the integral (\ref{eq:mmmm}) numerically for
  any $n>2$.}:
\begin{equation}
g_{mn}(\epsilon_1,\ldots,\epsilon_n,\bfkappa)=\half\int_{-1}^1 dx\,
\sum_{P_n} \prod_{i=1}^m u(x|\epsilon_{p_i},\bfkappa)\prod_{j=m+1}^n
[1-u(x|\epsilon_{p_j},\bfkappa)],
\label{eq:mmmm}
\end{equation}
where $P_n$ is the set of all permutations $(p_1,\ldots,p_n)$ of the numbers
$1,\ldots,n$, and $m\le n$. For example,
\begin{align}
g_{23}(\epsilon_1,\epsilon_2,\epsilon_3,\bfkappa)&= g_{22}(\epsilon_1,\epsilon_2,\bfkappa) 
+ g_{22}(\epsilon_2,\epsilon_3,\bfkappa)+g_{22}(\epsilon_3,\epsilon_1,\bfkappa)
- 3g_{33}(\epsilon_1,\epsilon_2,\epsilon_3,\bfkappa) \notag \\
g_{13}(\epsilon_1,\epsilon_2,\epsilon_3,\bfkappa)&= g_{11}(\epsilon_1,\bfkappa) + g_{11}(\epsilon_2,\bfkappa) + g_{11}(\epsilon_3,\bfkappa) -
2g_{22}(\epsilon_1,\epsilon_2,\bfkappa)-2g_{22}(\epsilon_2,\epsilon_3,\bfkappa) \notag \\
& \quad \quad \quad -2g_{22}(\epsilon_3,\epsilon_1,\bfkappa)
+3g_{33}(\epsilon_1,\epsilon_2,\epsilon_3,\bfkappa).
\label{eq:mmmmb}
\end{align}

The geometric selection matrix $G_{mn}$ (eq.\ \ref{eq:gdefqq}) is
simply $\langle
g_{mn}(R_\star/a_1,R_\star/a_2,\ldots,R_\star/a_l,\bfkappa)\rangle$,
the average of the geometric selection factor over the joint distribution
of stellar radius $R_\star$ and planetary semi-major axis $a$ for the
survey. To evaluate $G_{mn}(\bfkappa)$ we use the
separability assumption (\ref{eq:sep}) with respect to
$\epsilon=R_\star/a$. Thus 
\begin{equation}
G_{mn}(\bfkappa)=\int g_{mn}(\epsilon_1,\ldots,\epsilon_n,\bfkappa)\prod_{k=1}^n
f(\epsilon_k)d\log\epsilon_k,
\end{equation}
where $f(\epsilon)d\log\epsilon$ represents the probability
distribution of $\epsilon$ as modified by the survey selection effects. 

With this parametrization and equations (\ref{eq:one}) and (\ref{eq:mmmm}) it is
straightforward to show that $G_{mn}(\bfkappa)$ is given by the binomial distribution,
\begin{equation}
G_{mn}(\bfkappa)=\frac{n!}{2 m!(n-m)!}\int_{-1}^1 dx\,
U^m(x|\bfkappa)[1-U(x|\bfkappa)]^{n-m}=\half\int_{-1}^1 dx\,S_{mn}[U(x|\bfkappa)]
\label{eq:ggdef}
\end{equation}
where $S_{mn}$ is given by equation (\ref{eq:sdef}),
\begin{equation}
U(x|\bfkappa)\equiv \frac{\int f(\epsilon)u(x|\epsilon,\bfkappa)\,d\log\epsilon}{\int f(\epsilon)\,d\log\epsilon}=\sum_{\ell=0}^\infty
    Q_\ell(\bfkappa) B_\ell P_\ell(x),
\label{eq:udef}
\end{equation}
and
\begin{equation}
B_\ell\equiv \int f(\epsilon) b_\ell(\epsilon)\,d\log\epsilon\quad\mbox{with}\quad \int f(\epsilon)\,d\log\epsilon=1.
\label{eq:bldef}
\end{equation}
Since $B_\ell$ does not depend on the unknown parameters $\bfkappa$ of
the inclination distribution it can be evaluated once and for all at
the start of any optimization procedure.  It is straightforward to
show that the relations (\ref{eq:G}) are satisfied by these formulae,
and that the matrices $\mathbf{G}$ and $\mathbf{S}$ commute. In
numerical work we typically truncate infinite series such as
(\ref{eq:udef}) at $\ell=\ell_{\rm max}=50$, but for very thin disks
it may be necessary to include terms of higher $\ell$.

We pointed out in equation (\ref{eq:seprv}) that most survey selection
effects preserve the separability assumption. This result does not
generally hold for geometric selection effects. To illustrate this,
consider the simple case of a population of stars containing two
planets, with zero relative inclination. Write the probability distribution
of $\epsilon=R_\star/a$ of two-planet systems as
$f(\epsilon_1)f(\epsilon_2)d\log\epsilon_1d\log\epsilon_2$ (after
survey selection effects but before geometric selection effects). Then
using equation (\ref{eq:gtwotwo}) it is evident that the probability
distribution of two-tranet systems is
\begin{equation}
dp_2(\epsilon_1,\epsilon_2)=f(\epsilon_1)f(\epsilon_2)\hbox{min\,}(\epsilon_1,\epsilon_2)\,d\log\epsilon_1\,d\log\epsilon_2,
\end{equation}
which is not separable. Only for isotropic distributions do geometric
selection effects preserve separability.

\subsection{The inclination distribution}
\label{sec:fisher}

\noindent
In this paper we model the probability distribution of the inclinations
$dp=q(i|\kappa)di$ as a Fisher distribution, 
\begin{equation}
q(i|\kappa)=\frac{\kappa}{2\sinh\kappa}\exp(\kappa\cos i)\sin i.
\label{eq:idistf}
\end{equation} 
The parameter $\kappa$ is related to the mean-square value of $\sin i$
through
\begin{equation} 
\langle\sin^2i\rangle=\int
di\sin^2i\,q(i|\kappa)= 2\frac{\coth \kappa}{\kappa}-\frac{2}{\kappa^2}.  
\label{eq:rms}
\end{equation}
When $\kappa\ll1$ the Fisher distribution approaches an isotropic
distribution, $\lim_{\kappa\to0}q(i|\kappa)=\frac{1}{2}\sin i$, while
for $\kappa\gg1$ it approaches the Rayleigh distribution,
$\lim_{\kappa\to\infty}q(i|\kappa)=(2i/s^2)\*\exp(-i^2/s^2)$ where
$s=(2/\kappa)^{1/2}$ is the rms inclination and
$\frac{1}{2}\pi^{1/2}s=0.8862s$ is the mean inclination. The Rayleigh
distribution is commonly used to model the inclination distribution of
asteroids, Kuiper-belt objects, stars in the Galactic disk (where it
is known as the Schwarzschild distribution), etc. As
$\kappa\to-\infty$ the Fisher distribution approaches a retrograde
Rayleigh distribution.

For the Fisher distribution, equation (\ref{eq:onetwo}) becomes 
\begin{equation}
Q_\ell(\kappa)=\sqrt\frac{\pi\kappa}{2}\frac{I_{\ell+1/2}(\kappa)}{\sinh\kappa}
\end{equation}
where $I$ denotes a modified Bessel function. 

\subsection{Validity of the separability assumption}
\label{sec:valid}

\noindent
There is limited evidence on the accuracy of the separability 
approximation for multi-planet systems.  First consider RV surveys, in
which there are no geometric selection effects. The most important
survey selection effects depend only on the properties of an
individual planet so an RV survey of a separable parent distribution
should lead to a separable detected distribution (eq.\
\ref{eq:seprv}).

\cite{wri09} compare 28 multi-planet systems and a much larger number
of single-planet systems detected by RV surveys. They find that (i)
the eccentricities in multi-planet systems are smaller (mean
eccentricity 0.22, compared to 0.30 in single-planet systems); (ii)
the logarithmic semi-major axis distribution in multi-planet systems
is flatter, without the pileup of hot Jupiters between $0.03\au$ and
$0.07\au$ and the enhancement outside $1\au$ that are seen in
single-planet systems; (iii) multi-planet systems exhibit an
overabundance of planets with minimum mass between 0.01 and 0.2
Jupiter masses. These differences are incompatible with separability
and statistically significant ($p<0.03$), but relatively small: they
represent maximum differences of only 0.18, 0.17, and 0.26 in the
cumulative probability distributions for eccentricity, semi-major
axis, and minimum mass. \cite{wri09} point out that the last of these
differences may also be amplified by an unmodeled survey selection
effect---stars hosting planets tend to be observed more frequently,
thereby enhancing the chance to discover additional low-mass
planets. Most of the plots in the lower panels of Figure \ref{fig:one}
are marginally consistent with separability, as discussed at the
end of \S\ref{sec:survey}. 

The evidence on separability from the Kepler survey is more difficult to
interpret, because geometric selection effects do not 
preserve separability (see discussion just before
\S\ref{sec:fisher}). Nevertheless, the semi-major
axis distributions of single- and multiple-tranet systems in the Kepler survey
are indistinguishable according to a KS test ($p$-value 0.20;
see also Figure \ref{fig:epsw}), which is consistent with separability. Presumably the pileup of hot Jupiters
at small semi-major axes seen in the RV surveys is less prominent in
the Kepler sample because the typical planetary mass is much smaller,
and the jump outside $1\au$ is not seen because Kepler is not
sensitive to these orbital periods. 

\cite{latham11} have shown that Kepler systems with multiple tranets
are less likely to include a giant planet (larger than Neptune) than
systems with a single tranet. We confirm using a KS test that the
distributions of radii in the single- and multiple-tranet systems are
different (maximum difference in the cumulative probability
distribution of 0.20). However, the results at the end of \S\ref{sec:survey} show
that the numbers of two-, three-, and four-tranet systems as a
function of the radius cutoff appear to be consistent with
separability. Evidently equations such as (\ref{eq:cccc}) that we use
to compare the observable multiplicity function between surveys are less
sensitive to deviations from separability than statistical tests
designed specifically for this purpose.

These comparisons suggest that deviations from separability,
though present in both the RV and Kepler planet samples, 
are not large enough to compromise our method and results. However, further
exploration of both the magnitude and the effects of these deviations is needed. 

\section{Estimating the inclination distribution and the
  multiplicity function from the Kepler survey}

\label{sec:kepler}

\subsection{Properties of the survey}

\noindent
The Kepler survey has a complex set of survey selection effects, which
we do not attempt to model. The constraints on the multiplicity
function that we derive therefore apply to the population of planets
in radius, semi-major axis, etc.\ that Kepler detects, whatever that
population may be (for a discussion of selection effects and
completeness in the Kepler catalog see \citealt{howard} and
\citealt{youdin}). If we denote the multiplicity function of this
population by $\mathbf{N}$ and the observable multiplicity function of
the Kepler survey by $\overline{\mathbf{n}}$ then equation
(\ref{eq:gdefqq}) becomes
\begin{equation}
\overline{\mathbf{n}}=\mathbf{G}\cdot\mathbf{N}.
\label{eq:gdeftwo}
\end{equation}
The validity of this equation requires only the plausible assumption
that the probability that Kepler will detect a given transiting planet
around a given star is independent of whether it detects other
transits around the same star.

To produce a more homogeneous sample, we trim the catalog of
\cite{bor11} to include only stars with effective temperatures between
4000 and 6500 K and surface gravity $\log g>4.0$ (roughly equivalent
to FGK dwarfs), and to Kepler magnitudes between 9.0 and 16.0; this
trimming leaves 124,613 stars from the original sample of 153,196. We
also restrict the catalog to planets with orbital period less than 200
d and radius less than 2 Jupiter radii; this leaves 1092 planets from
the original sample of 1235. The numbers of stars with $0,1,2,\ldots$ 
tranets are
\begin{equation}
\begin{array}{cccc}n_0=1.237\times 10^5,\quad & n_1=737, \quad&  n_2=104, \quad&
n_3=37, \\ n_4=7, \quad& n_5=1, \quad& n_6=1, & n_k=0\hbox{ for
}k>6.\end{array}
\label{eq:plnum}
\end{equation}

\begin{figure}
\centering
\includegraphics[clip=true,width=0.95\hsize]{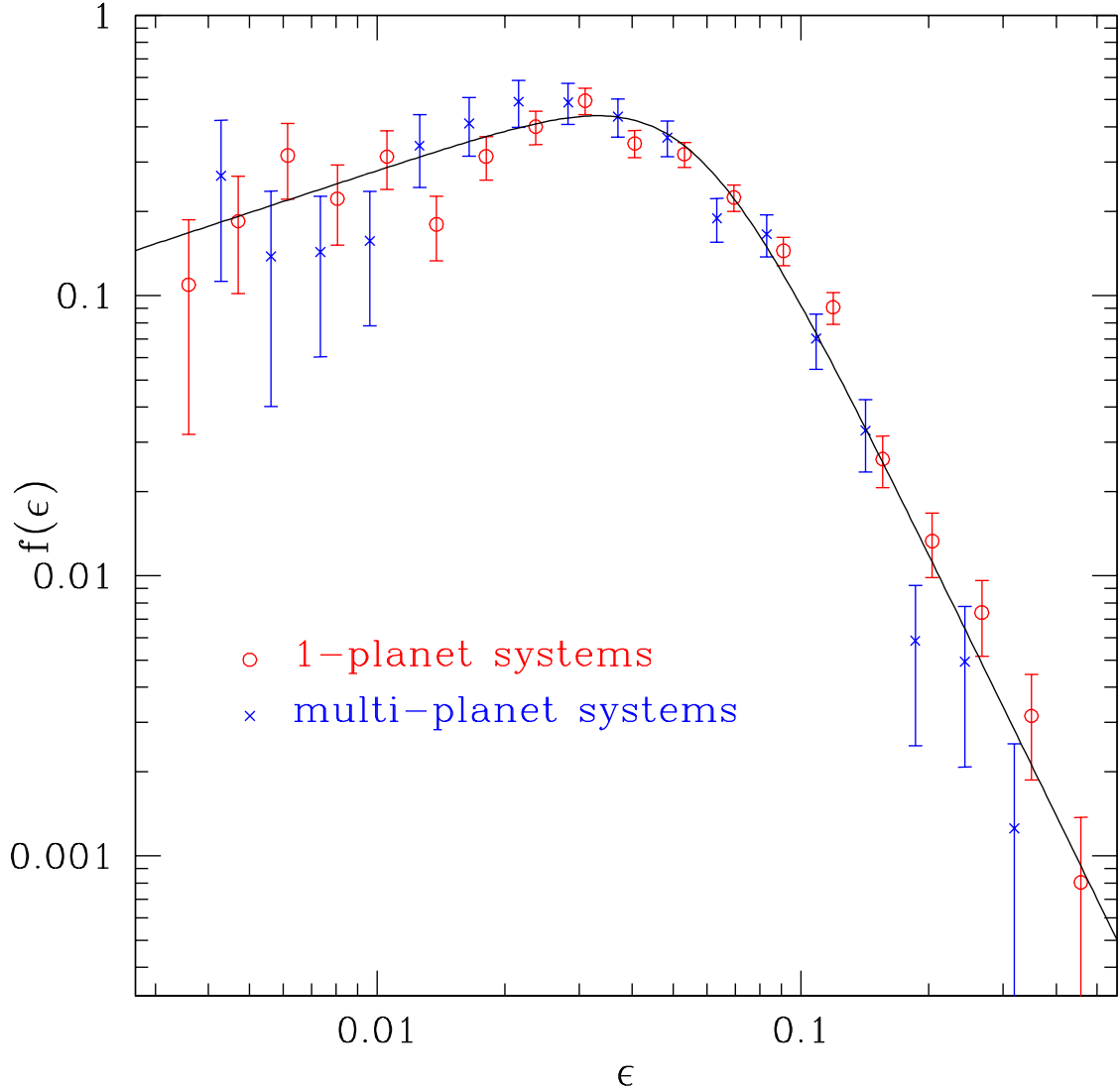}
\caption{The probability distribution of $\epsilon=R_\star/a$, the
  ratio of stellar radius to planetary semi-major axis, for tranets
  detected by Kepler. The differential probability distribution is
  $dp=f(\epsilon)\,d\log\epsilon$. The data points for single- and
  multi-tranet systems are shown separately. The solid line shows the
  analytic fitting formula (\ref{eq:ffit}).}  \label{fig:epsw}
\end{figure}

We need to determine the function $f(\epsilon)$, where
$f(\epsilon)\,d\log\epsilon$ is the fraction of planets in the range
$d\log\epsilon$ given the intrinsic distribution of planets and the
survey selection effects for Kepler. As usual $\epsilon=R_\star/a$ is
the ratio of stellar radius to planetary semi-major axis; the stellar radius is determined from the
host-star mass and surface gravity and the semi-major axis is
determined from the host-star mass and the planetary orbital
period. Figure \ref{fig:epsw} shows data points for 
$f(\epsilon)$ from single-tranet systems (red points) and from
planets in multi-tranet systems (blue points). The
data points have been constructed by adding a contribution of
$\epsilon^{-1}$ (to account for geometric selection effects) from each
tranet to the corresponding bin, then normalizing so that the integral over $\log\epsilon$ is unity. The
distributions for single-tranet and multi-tranet systems are quite
similar, and can be adequately fit by the parametrization
\begin{equation}
f(\epsilon) =
0.656\frac{(\epsilon/\epsilon_0)^{0.5}}{1+(\epsilon/\epsilon_0)^{3.6}},
\quad \epsilon_0=0.055,\quad\mbox{for }\epsilon>0.004
\label{eq:ffit}
\end{equation}
and zero for $\epsilon < 0.004$. The sharp decline for
$\epsilon\gtrsim0.1$ is due to an absence of planets with semi-major
axis $\lesssim0.04\,\textsc{au}$ \citep{bor11}, while the cutoff at
$\epsilon\lesssim0.004$ is due to the limited timespan of the Kepler
data. 

\subsection{Statistical method}

\noindent
The probability that the survey actually detects
$\{n_0,n_1,\ldots,n_K\}$ stars having $0,1,\ldots,K$ planets
is
\begin{equation}
P(\mathbf{n}|\mathbf{N},\bfkappa)=\prod_{k=0}^K \frac{\overline{n}_k^{n_k}\exp(-\overline{n}_k)}{n_k!}
\label{eq:likeone}
\end{equation}
where $\mathbf{n}=(n_0,n_1,\ldots,n_K)$ and $\overline{n}_k$ is related to
$\mathbf{N}$ by equation (\ref{eq:gdeftwo}). 

Estimating the multiplicity function $\mathbf{N}$ and the
inclination distribution parameters $\bfkappa$ from $\mathbf{n}$ is a
straightforward but challenging problem in statistics and
optimization. This problem can be attacked with a variety of
methods (linear programming, minimum $\chi^2$, maximum
likelihood, Bayesian analysis using a Markov chain Monte Carlo
algorithm, etc.), and we have experimented with most of these. In this paper
we have usually chosen maximum likelihood, as a reasonable compromise between
generality, computation time, and clarity of interpretation.

The log of the likelihood of a given observational result $\mathbf{n}$
is
\begin{equation}
\log P(\mathbf{n}|\mathbf{N},\bfkappa)=\sum_{k=0}^K
n_k\log\left[\sum_{l=k}^K
  G_{kl}(\bfkappa)N_l\right]-\sum_{k=0}^K\sum_{l=k}^K
  G_{kl}(\bfkappa)N_l-\sum_{k=0}^K\log n_k!.
\label{eq:liketwo}
\end{equation}
Note that the second term on the right can be simplified to
$\sum_lN_l$ using equation (\ref{eq:sumG}). We then
maximize $\log P$ with respect to $\mathbf{N}$ and
$\bfkappa$, subject to the constraint $N_k\ge0$,
$k=0,\ldots,K$.

\subsection{Results}

\begin{figure}
\centering
\begin{tabular}{c}
\includegraphics[clip=true,width=0.52\hsize]{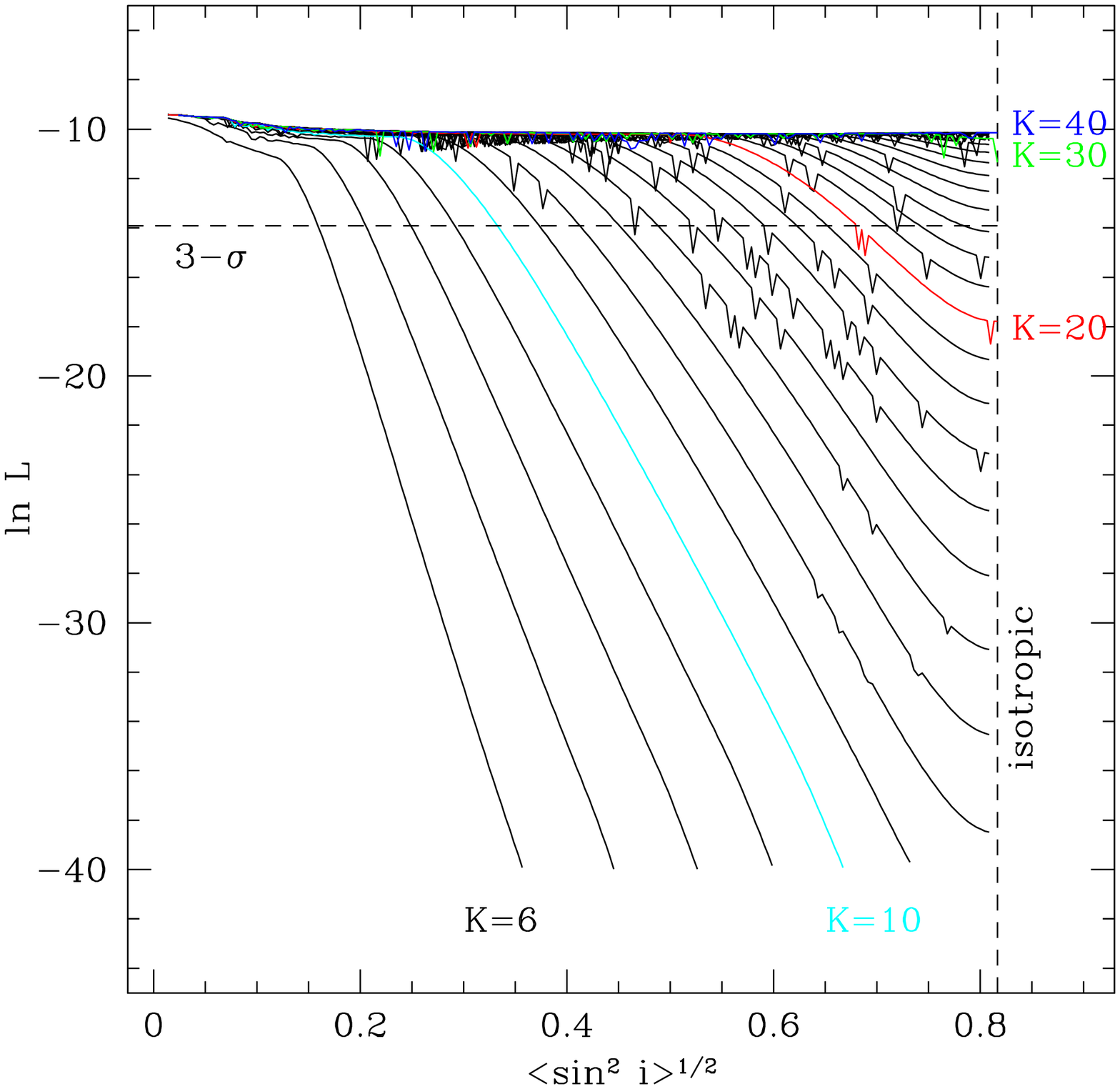} \\
\includegraphics[clip=true,width=0.52\hsize]{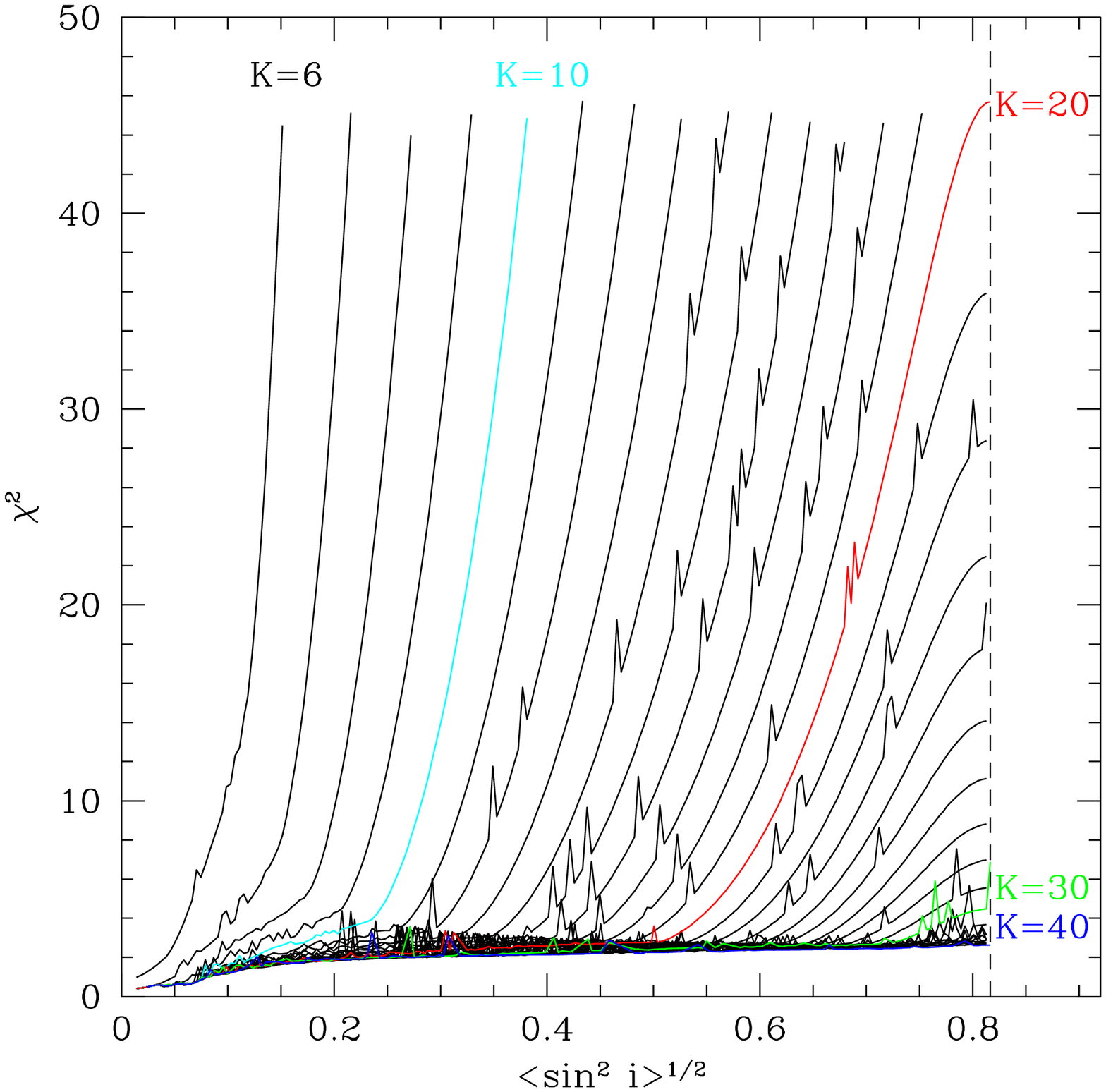} 
\end{tabular}
\caption{(top) The maximum likelihood of solutions for the
  multiplicity function of the Kepler survey, as a function of rms
  inclination and maximum number $K$ of planets per system. Solid
  lines connect solutions with a given $K$, $6\le K\le 40$; lines for
  $K=10,20,30,40$ are colored cyan, red, green, and blue for
  emphasis. The vertical dashed line denotes isotropic planetary
  systems. The horizontal dashed line marks
  systems that are 3-$\sigma$ ($\Delta\ln L=4.5$) lower in likelihood
  than razor-thin solutions, $\langle\sin^2i\rangle=0$. (bottom) Plots of $\chi^2$ (eq.\
  \ref{eq:chisq}) for the maximum-likelihood models shown above. We
  estimate that models with $\chi^2\lesssim 5$ are good fits to the
  data.}
\label{fig:maxinc}
\end{figure}

\noindent
The top panel of Figure \ref{fig:maxinc} shows the maximum likelihood
as a function of the rms inclination and the maximum number of planets
per system, $K$, for $6\le K\le 40$.  The minimum allowed value is
$K=6$ since Kepler has found one system with six tranets. The
maximum-likelihood models with a given $K$ are connected to form solid
lines, and the families with $K=10$, 20, 30, and 40 are colored for
emphasis. There are occasional small dips in the lines when the
optimization algorithm (a quasi-Newton algorithm from NAG)  converged
on a local rather than global maximum. The figure shows that:

\noindent (i) The highest likelihood is for razor-thin systems, with near zero rms
inclination. However, the preference for zero rms inclination has only
marginal statistical significance: systems exist at all rms
inclinations---even isotropic systems---with log likelihood only 0.73
smaller than the razor-thin solutions.

\noindent (ii) Systems with large rms inclinations are only consistent
with the data if a fraction of them contain a large number of
planets. At the 3-$\sigma$ level (log likelihood smaller than the
maximum by 4.5, marked by a horizontal dashed line on the figure), the
maximum rms inclination is related to the maximum number of planets by
\begin{equation}
\langle\sin^2i\rangle^{1/2}\le\left\{\begin{array}{ll}0.15+0.037(K-6), &
    K<24 \\ (\frac{2}{3})^{1/2}\mbox{ (isotropic)}, & K\ge
    24. \end{array}\right.
\label{eq:maxp}
\end{equation}

It is possible, of course, that even the maximum-likelihood model does
not fit the data well. To explore this possibility, we have calculated
the standard Pearson $\chi^2$ statistic,
\begin{equation}
\chi^2=\sum_{k=0}^K\frac{(n_k-\overline{n}_k)^2}{\overline{n}_k}=\sum_{k=0}^K\frac{(n_k-\sum_l
  G_{kl}N_l)^2}{\sum_l G_{kl}N_l}. 
\label{eq:chisq}
\end{equation}
The distribution of the $\chi^2$ statistic is not straightforward to
interpret, since $\overline{n}_k\lesssim 1$ for many $k$ and since the
number of degrees of freedom is not well-defined. Nevertheless it
is probably reasonable to expect that there is a good fit to the data
if $\chi^2\lesssim 5$. The values of $\chi^2$ for the
maximum-likelihood solutions in the top panel of Figure
\ref{fig:maxinc} are shown in the bottom panel of that figure. There
are satisfactory models with all rms inclinations, but as before such
models require that some systems contain many planets if the rms inclination is large.

It is instructive to examine the isotropic solution with $K=30$ in
more detail (the behavior of the isotropic solutions with $K>30$ is
qualitatively similar). The fraction of stars with $k$-planet systems is
\begin{equation}
\frac{N_k}{\sum_{l=0}^K
  N_l}=\left\{\begin{array}{ll}0.944 & k=0, \\ 0.0065 & k=1, \\ 0 &
    k=2, \\ 0.0452 & k=3, \\ 0 &
      k=4,\ldots,29, \\ 0.0043 & k=30.
     \end{array}\right.
\end{equation}
Thus, in this solution, about half of the planets are contained in
three-planet systems, and the other half in a small population
($<0.5\%$) of stars with many-planet systems. This multiplicity
function and inclination distribution are neither unique nor
particularly plausible but they are consistent with the Kepler data. 

\begin{figure}
\centering
\includegraphics[clip=true,width=0.95\hsize]{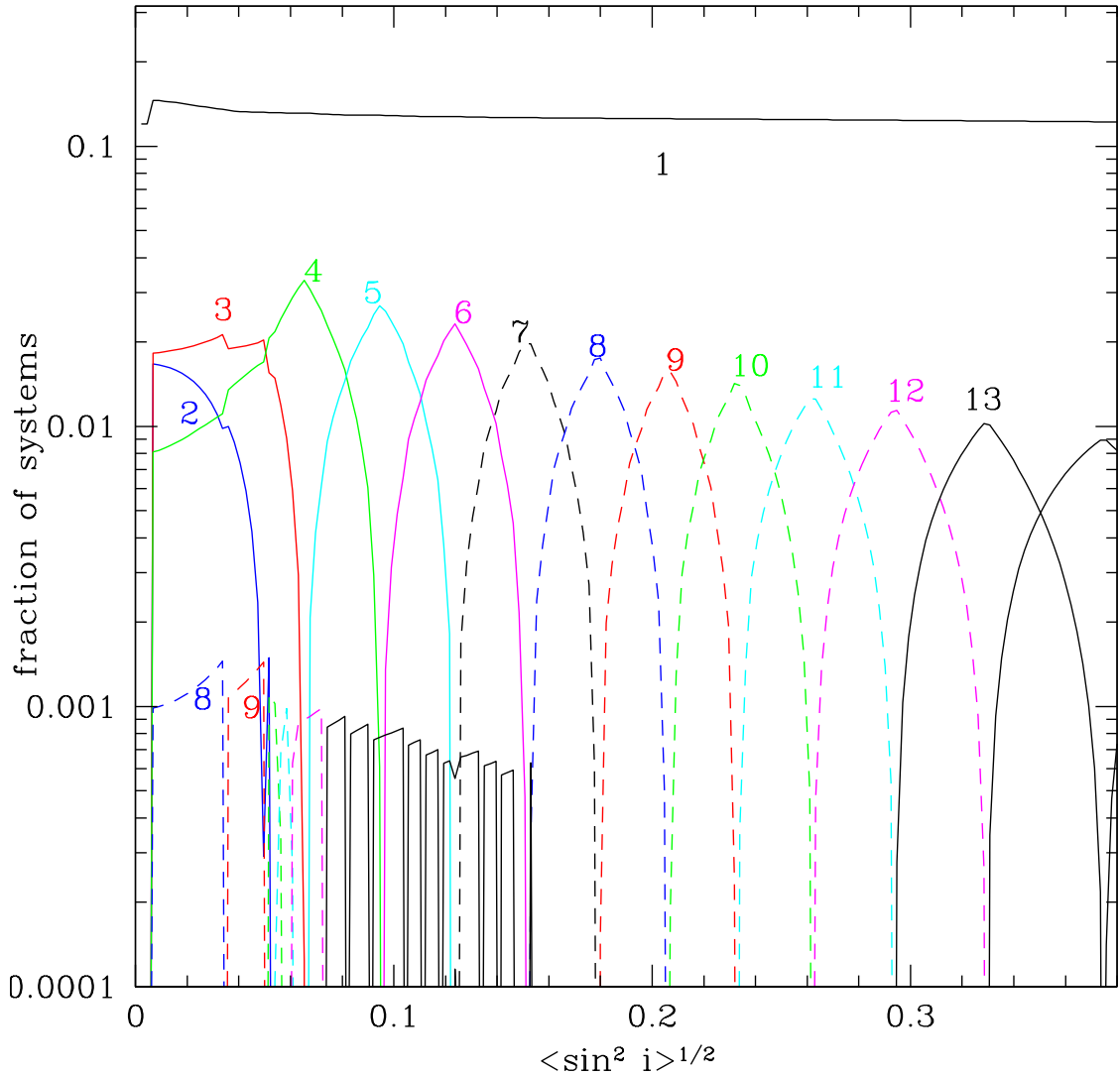}
\caption{The fraction of stars in the Kepler sample containing
  $k$-planet systems, as a function of the rms value of
  $\sin i$. The curves are labeled by $k$ for $k\le 13$ and curves
  with $7 \le k\le 12$ are dashed. These curves were obtained by
  linear programming, using the constraint that $\overline{n}_k$ must
  lie within the 90\% confidence interval determined through
  equation (\ref{eq:likeone}). The cost function minimized the total
  number of planets but the result is insensitive to this choice.} \label{fig:npl}
\end{figure}

Figure \ref{fig:npl} shows the fraction of stars in the Kepler sample
with $0,1,2,3,\dots$-planet systems, as a function of the assumed rms
inclination. The results are for $K=30$ but are qualitatively similar
for larger values of $K$. Our initial attempts to construct this figure
were unsuccessful, because the appearance of the figure is very
sensitive to cases when the optimization algorithm settles on a local
maximum of the likelihood. To avoid this difficulty, we re-cast the
optimization as a problem in linear programming: we demanded that each
$\overline{n}_k$ should lie within the 90\% confidence interval
determined by the Poisson distribution (\ref{eq:likeone}), and from
these solutions we chose the one with the minimum total number of
planets $\sum_{k=1}^KN_k$. This specifies a unique
solution, if one exists.

At the smallest inclinations ($\langle \sin^2 i\rangle^{1/2} < 0.05$)
the solution contains a mix of 1,2,3,4, and 8 or 9-planet systems. As the
rms inclination increases, the mixture becomes strongly dominated by
1-planet and $n_i$-planet systems where $n_i$ varies monotonically
with the rms inclination---for example, $n_i=12$ when $\langle \sin^2
i\rangle^{1/2} \simeq 0.3$. We caution that these results should not be
regarded as a prediction of the Kepler multiplicity function for a
given rms inclination.

The need for many-planet systems is straightforward to
understand. Consider the extreme case of an isotropic
distribution. Then $\kappa=0$ and $q(i|\kappa=0)=\frac{1}{2}\sin i$; thus
$Q_{\ell}(\kappa=0)=\delta_{\ell 0}$ from equation (\ref{eq:onetwo})
and the orthogonality properties of the Legendre polynomials. Thus
$U(x|\kappa=0)=B_0$ (eq.\ \ref{eq:udef}) and using equation
(\ref{eq:ggdef})
\begin{equation}
G_{mn}(\bfkappa)=\frac{n!}{m!(n-m)!}B_0^m(1-B_0)^{n-m}.
\label{eq:ggdefiso}
\end{equation}
If all systems contain $n$ planets, the ratio of the number of
$m$-tranet systems to the number of $(m+1)$-tranet systems is
\begin{equation}
  \frac{G_{mn}}{G_{m+1,n}}=\frac{m+1}{n-m}\frac{1-B_0}{B_0}, \quad
    n\ge m+1.
\label{eq:grat}
\end{equation}
Using equations (\ref{eq:bldef}) and (\ref{eq:ffit}) we find that
$B_0=0.0321$ for the Kepler survey. From equation
(\ref{eq:plnum}) we find $n_1/n_2=7.1\pm0.7$. For comparison the ratio
$G_{1n}/G_{2n}$ is less than $7.1+0.7=7.8$ only for $n\ge 9$; thus
any population dominated by systems with less than 9 planets will overproduce
1-tranet systems relative to 2-tranet systems. Similarly, for the
Kepler survey $n_2/n_3=2.8\pm0.5$, and $G_{2n}/G_{3n}>2.8+0.5=3.3$ unless
$n\ge 30$. 

\begin{figure}
\centering
\includegraphics[clip=true,width=0.95\hsize]{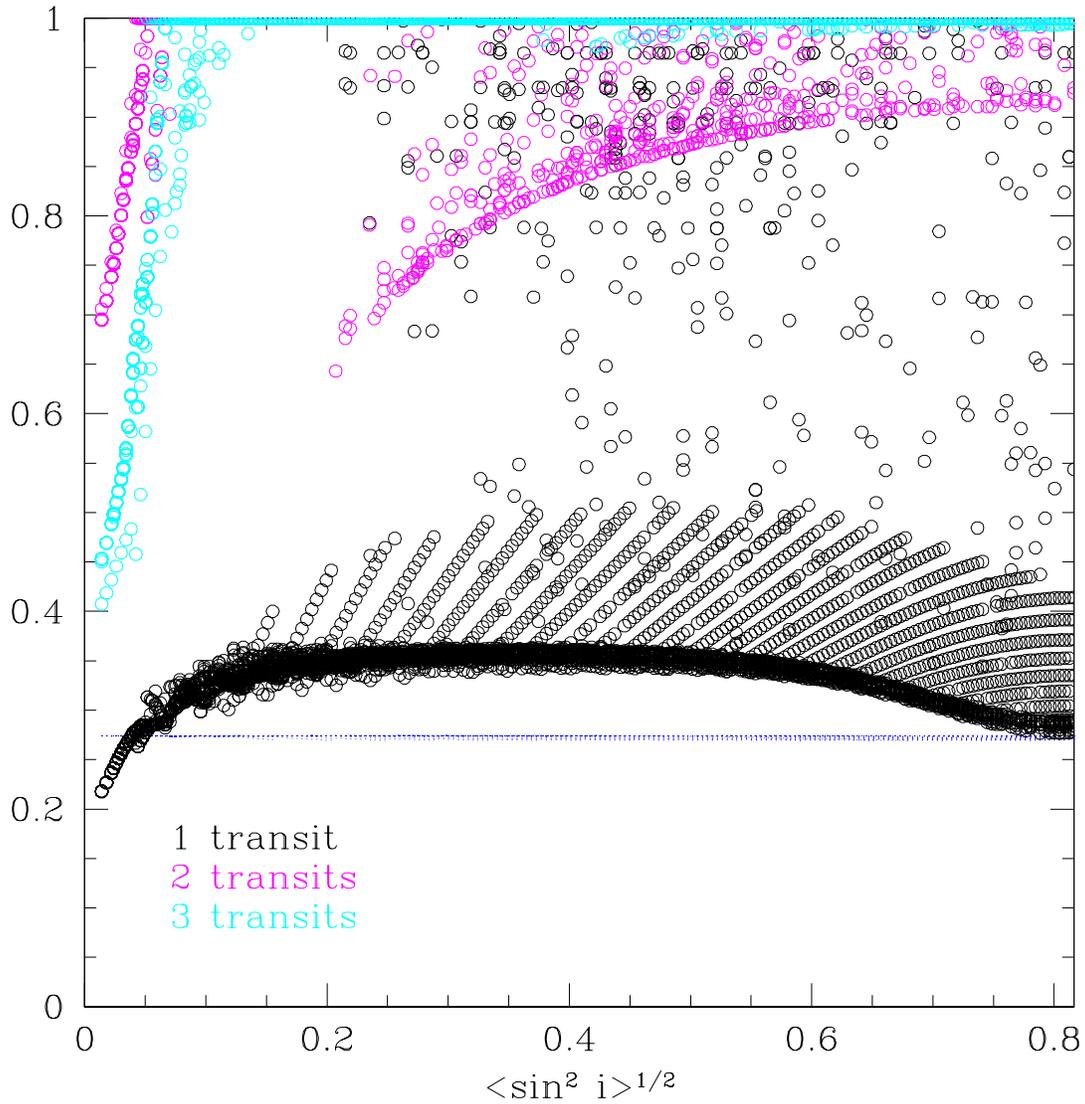}
\caption{The horizontal blue line, composed of $\sim 7000$ points from
  individual maximum-likelihood models, shows the average
  number of planets per star in the Kepler sample, as a function of
  the rms inclination and the maximum number of planets per star,
  $11\le K\le 40$. The large open circles show the probability that a
  system exhibiting one, two, or three tranets has additional
  planets. } \label{fig:total}
\end{figure}

The average number of planets per star from these solutions is shown
in Figure \ref{fig:total}. This result is insensitive
to the rms inclination and the maximum number of planets per star
($K$), since it is given simply by the ratio of the total number
of planets to the number of target stars, divided by the probability
that a single randomly oriented planet will transit \citep{youdin}. Mathematically,
\begin{equation}
\langle\mbox{number of planets per star}\rangle=\frac{\sum_{k=1}^K
  kn_k}{B_0\sum_{k=0}^Kn_k}=0.274.
\end{equation}

The large open circles in Figure \ref{fig:total} show the probability
that a system with one, two, or three tranets has additional planets. Typically the
fraction of one-tranet systems with additional planets is 0.2--0.5,
without a strong dependence on rms inclination. For two or three
tranets the probability that there are additional unseen planets is
substantially higher. The additional planets may be detectable by
transit timing variations \citep{ford11}. 

\section{Combining Kepler and radial-velocity surveys}

\label{sec:keprv}

\begin{figure}
\centering
\includegraphics[clip=true,width=0.95\hsize]{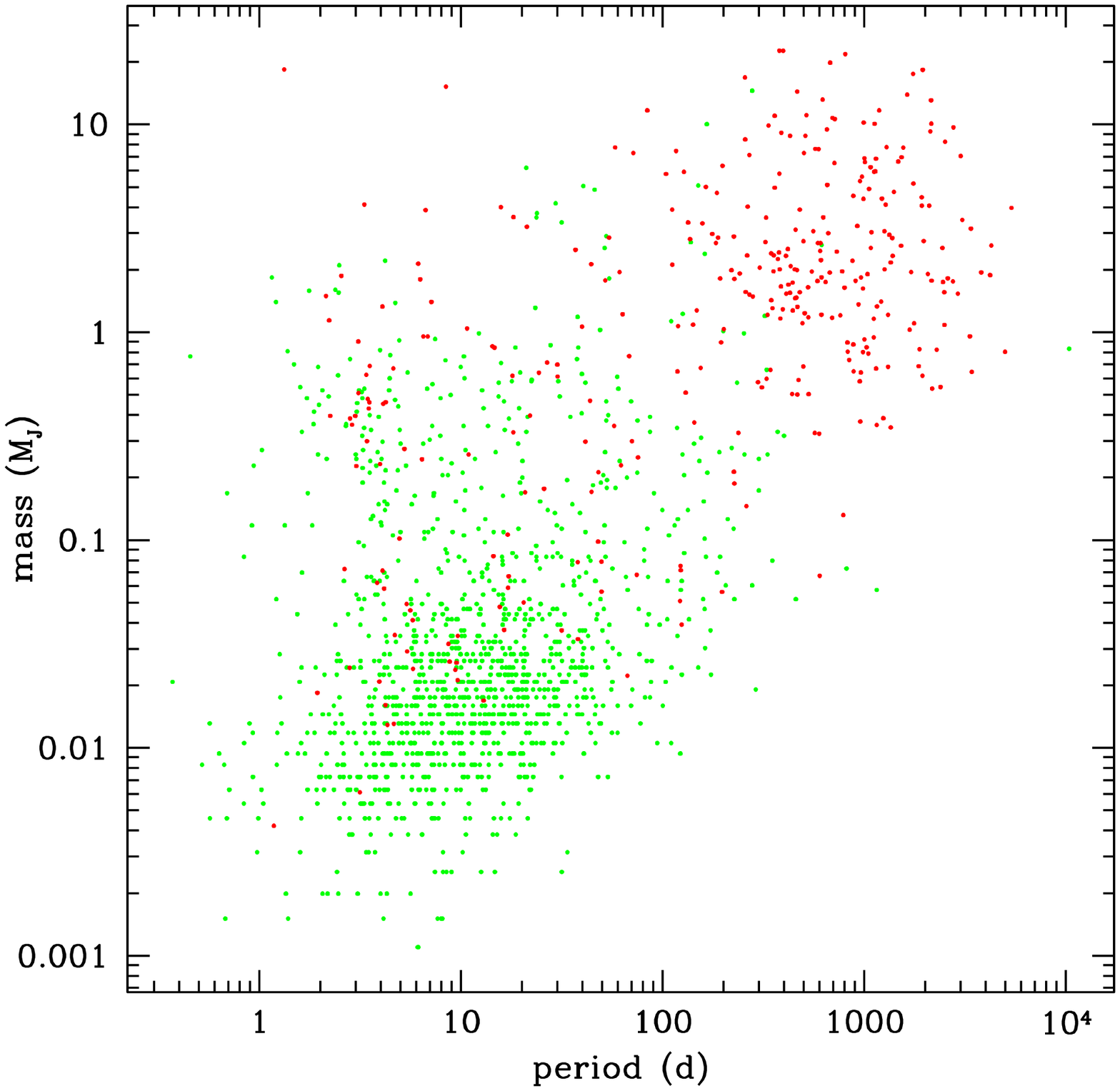}
\caption{The orbital periods and masses of the planets detected by
  Kepler (green) and by ground-based radial-velocity surveys
  (red). Orbital periods are in days and masses are in Jupiter
  masses. Masses $M$ for transiting planets are computed from radii
  $R$ using $M=(R/R_\oplus)^{2.06}M_\oplus$ \citep[][for a more
  accurate relation see eq.\ \ref{eq:mr}]{liss11b}
  and masses for radial-velocity planets are minimum masses $M\sin
  \gamma$.} \label{fig:trv}
\end{figure}

\noindent
As described in the Introduction, a comparison of the observable multiplicity
functions of planetary systems detected by radial velocities and by
transits can offer a powerful probe of the inclination
distribution. The principal obstacle to making this comparison is that
the masses and orbital periods of the planets detected through these
two observational techniques are quite different, as illustrated in
Figure \ref{fig:trv}, and the multiplicity functions in these
two regions of parameter space are likely to be different. In this
section we use the separability approximation and the methods of
\S\ref{sec:survey} to overcome this obstacle. 

Suppose that we wish to combine the Kepler survey with a
radial-velocity (RV) survey (or a set of such surveys). The surveys
yield $n_k^\mathrm{Kep}$ and $n_k^\mathrm{RV}$ systems containing $k$
planets. We assume that both surveys have similar target star
populations (we cull the list of target stars in both cases to include
only FGK dwarfs), with multiplicity function $\mathbf{N}$ for
Kepler and $c\,\mathbf{N}$ for the RV survey, where $c<1$ is a constant to
be determined. Let $\mathbf{S}(W^\mathrm{Kep})$ and
$\mathbf{S}(W^\mathrm{RV})$ be the survey selection functions. We
assume that there are no geometric selection effects for the RV
surveys (cf.\ footnote 2). The generalization of equation
(\ref{eq:likeone}) for the likelihood is
\begin{equation}
P(\mathbf{n}^\mathrm{Kep},\mathbf{n}^\mathrm{RV}|\mathbf{N},\bfkappa)=\prod_{k=0}^K\frac{(\overline{n}_k^\mathrm{Kep})^{n_k^\mathrm{Kep}}\exp(-\overline{n}_k^\mathrm{Kep})}{ n_k^\mathrm{Kep}!}\prod_{k=1}^K\frac{( \overline{n}_k^\mathrm{RV})^{n_k^\mathrm{RV}}\exp(-\overline{n}_k^\mathrm{RV})}{n_k^\mathrm{RV}!}
\label{eq:likefive}
\end{equation}
where
\begin{equation}
 \overline{\mathbf{n}}^\mathrm{Kep}=\mathbf{G}(\bfkappa)\mathbf{S}(W^\mathrm{Kep})\mathbf{N}, \quad
  \overline{\mathbf{n}}^\mathrm{RV}=c\,\mathbf{S}(W^\mathrm{RV}).
\label{eq:cccca}
\end{equation}
Notice that the second product in equation (\ref{eq:likefive}) starts
at $k=1$ since it is difficult to determine accurately how many stars
have been unsuccessfully examined for planets by RV methods (see
further discussion below). We then maximize the likelihood (\ref{eq:likefive}) over
$N_0,N_1,\ldots,N_K$, $W^\mathrm{Kep}$, $W^\mathrm{RV}$, and $c$ (as
shown in \S\ref{sec:survey}, the likelihood actually depends only on
the ratio $W^\mathrm{RV}/W^\mathrm{Kep}$). 

\begin{figure}\centering
\begin{tabular}{c}
\includegraphics[clip=true,width=0.58\hsize]{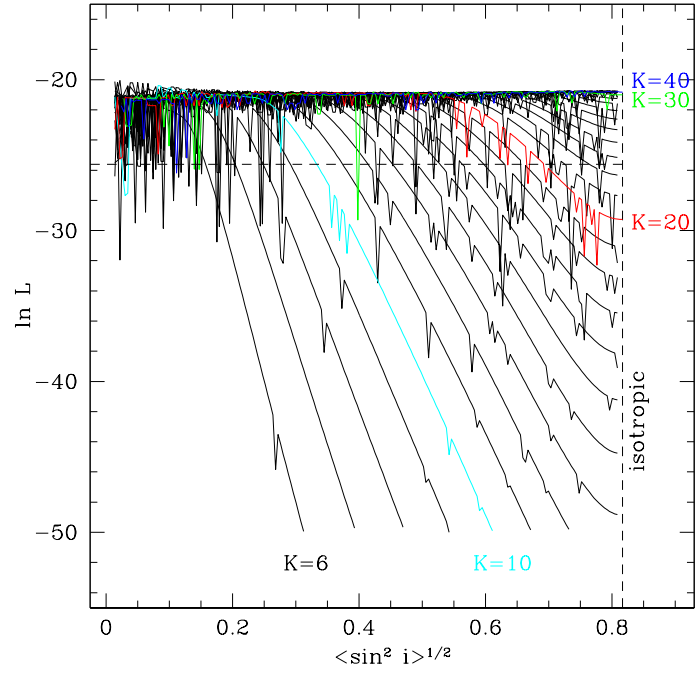} \\
\includegraphics[clip=true,width=0.58\hsize]{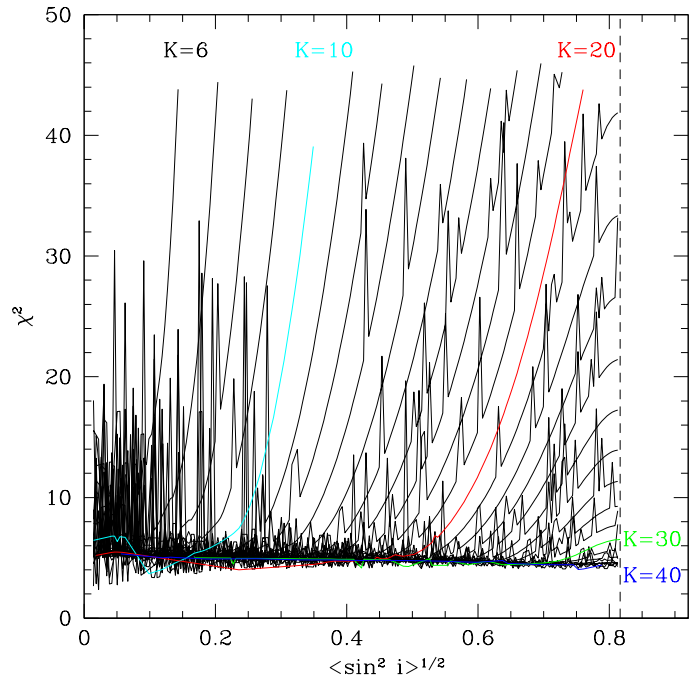} 
\end{tabular}
\caption{As in Figure \ref{fig:maxinc}, except the data include both the
  Kepler transit survey and radial-velocity surveys.}\label{fig:maxincrv}
\end{figure}

We determine the observable multiplicity function for RV planets using all
planets with FGK dwarf host stars in the exoplanets.org database \citep{wri11}
as of August 2010,
\begin{equation}
n_1^\mathrm{RV}=162, \quad n_2^\mathrm{RV}=24, \quad
n_3^\mathrm{RV}=7, \quad  n_4^\mathrm{RV}=1, \quad n_5^\mathrm{RV}=1,
\quad n_k^\mathrm{RV}=0\hbox{ for }k>5,
\label{eq:plnumrv}
\end{equation}
for a total of 240 planets.  The observable multiplicity function for Kepler
planets is given in equation (\ref{eq:plnum}). Figure
\ref{fig:maxincrv} shows the maximum likelihood as a function of the
rms inclination and the maximum number of planets per system, $K$
(top), as well as $\chi^2$ for these models (bottom). The plots are
noisier than Figure \ref{fig:maxinc}, presumably because the
optimization algorithm was less successful at finding the global
maximum likelihood, but otherwise look similar. In particular, systems
with large rms inclinations are consistent with the data if and only
if they contain a large number of planets. Evidently adding data from RV surveys has not significantly
tightened the constraints on the inclination distribution.

\begin{figure}\centering
\includegraphics[clip=true,width=0.95\hsize]{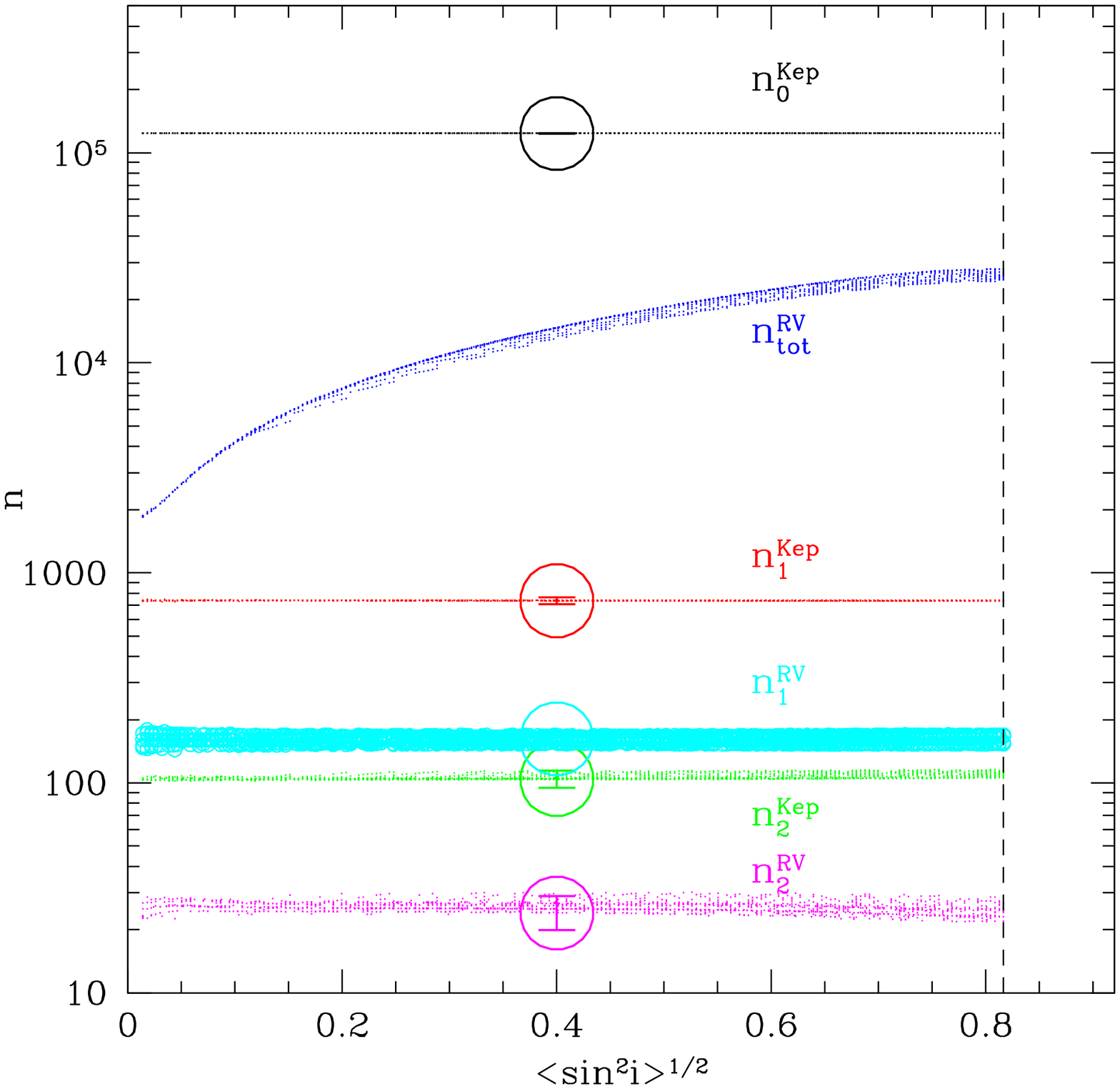}
\caption{The expected numbers of 0,1,2,3 tranet systems from the
  Kepler survey and of 1,2,3 planet systems from RV surveys, as
  predicted by our models. The observed numbers are shown as error
  bars surrounded by circles. Also shown is the total number of
  targets in the RV surveys as predicted by our models (blue points).}\label{fig:rv_test}
\end{figure}

We now show that adding information on the total number of target
stars in the RV surveys does allow the inclination distribution to be
determined. Figure \ref{fig:rv_test} shows the expected numbers
$\overline{n}_k^\mathrm{Kep}$ and $\overline{n}_k^\mathrm{RV}$ of
$k$-tranet systems from the Kepler survey and $k$-planet systems from
the RV surveys, as determined from the maximum-likelihood solutions
described above. Each point corresponds to a given maximum number of
planets ($6\le K \le 40$) and rms inclination, and only solutions
within 3--$\sigma$ of the global maximum likelihood are shown. The
points with error bars (surrounded by circles for greater visibility)
correspond to the observed numbers $n_k^\mathrm{Kep}$ and
$n_k^\mathrm{RV}$ from equations (\ref{eq:plnum}) and
(\ref{eq:plnumrv}). Most of the expected values lie within the error
bars of the corresponding observed value; this is no more than a
confirmation that our optimization code is performing properly. The
blue points show the total number of stars in the RV survey,
$\overline{n}_\mathrm{tot}^\mathrm{RV}=\sum_{k=0}^K
\overline{n}_k^\mathrm{RV}$, as determined by the optimization
code. The plot shows that $\overline{n}_\mathrm{tot}^\mathrm{RV}$ is
tightly correlated with the rms inclination, so an accurate
characterization of the total number of RV target stars would enable
the determination of the rms inclination.

This task is challenging given the heterogeneous surveys
that have produced the RV planets known at the present time. We have
used two distinct approaches, which we now describe.

\noindent (i) \cite{cum08} carry out a careful examination of
selection effects in the Keck Planet Search, and derive the percentage
of F, G, and K stars with a planet in various ranges of orbital period
and mass. The sample of RV planets used in our analysis (eq.\
\ref{eq:plnumrv}) is not corrected for selection effects, but for
sufficiently massive planets and sufficiently short orbital periods it
should be complete. For example, for planets more massive than
Jupiter, $M\sin\gamma>M_\mathrm{J}$, with orbital periods less than
one year, $P<1\mbox{ yr}$, the velocity semi-amplitude $K_\mathrm{RV}>30\,\mbox{m
  s}^{-1}$, large enough to be detectable in most surveys. In this
mass and period range our sample contains 46 planet-hosting stars and
\cite{cum08} estimate that the fraction of stars with planets is
$0.019\pm0.007$, which implies $n_\mathrm{tot}^\mathrm{RV}=2400\pm
900$. Altering the period range to $P<100\mbox{ d}$ gives
$n_\mathrm{tot}^\mathrm{RV}=2500\pm 1200$ (based on 21 host stars);
altering the mass cutoff to $M\sin\gamma>0.5M_\mathrm{J}$ gives
$n_\mathrm{tot}^\mathrm{RV}=1900\pm 500$ (based on 63 host
stars). This last estimate of $n_\mathrm{tot}^\mathrm{RV}$ is probably
low because the surveys we have used are not all complete at this
level.

\begin{figure}
\centering
\includegraphics[clip=true,width=0.95\hsize]{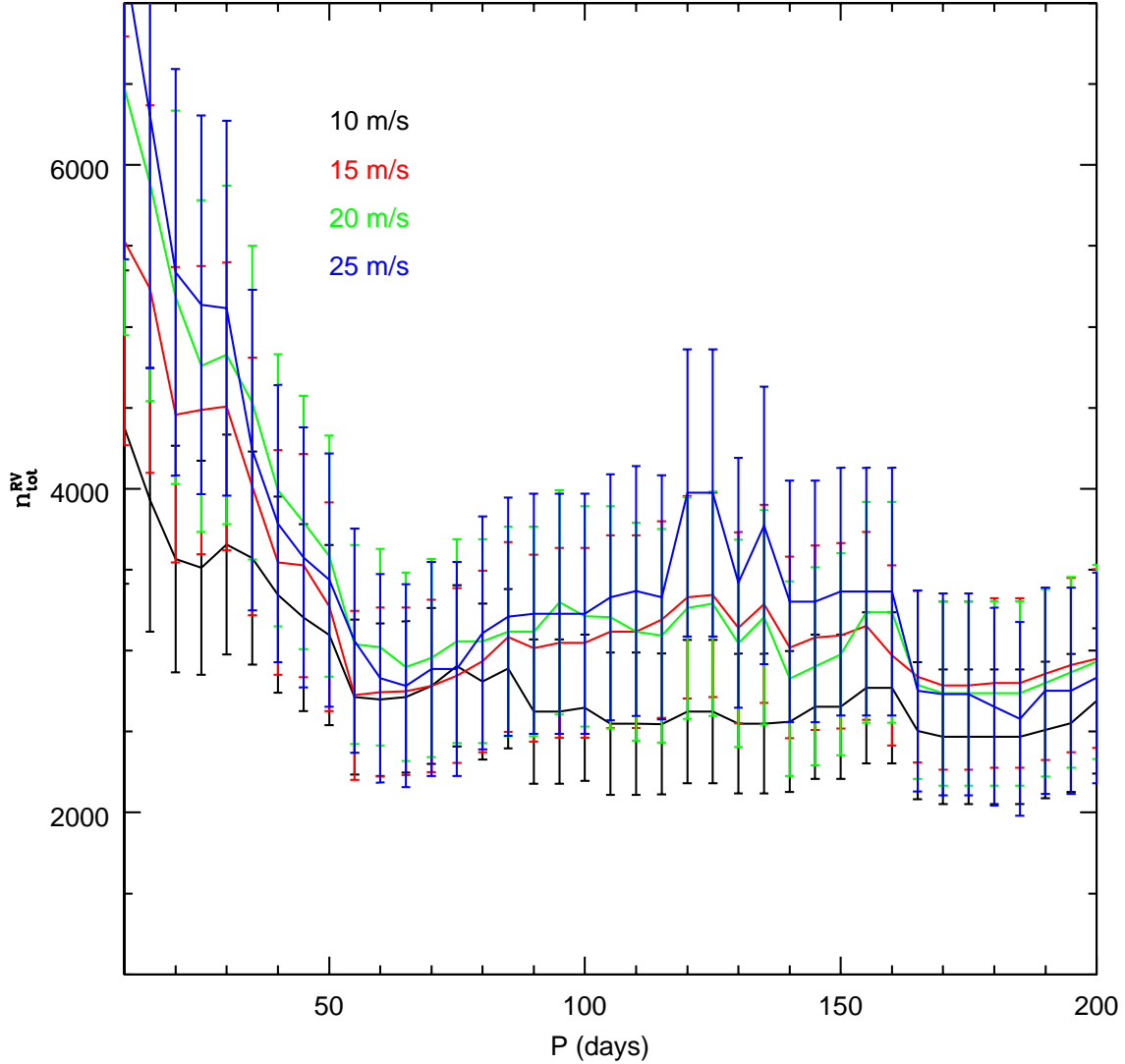}
\caption{The estimated number of host stars in RV surveys, as
  determined by comparison with the Kepler survey. The curves and
  associated error bars show the number of RV host stars as estimated
  by comparing the number of RV and Kepler planets with period less than $P$ and
  mass exceeding that required to induce a given velocity
  semi-amplitude $K_\mathrm{RV}$ at period $P$. The observed number of Kepler
  planets is multiplied by $g_{11}^{-1}$ (eq.\ \ref{eq:goneone}) to
  correct for geometric selection effects, and the conversion between
  radius and mass is given by equation (\ref{eq:mr}). Results are shown for
  four semi-amplitudes, $K_\mathrm{RV}=25,20,15,10\mbox{\,m s}^{-1}$; the plot at
  the smallest semi-amplitude is low because the RV surveys are
  incomplete at this level.} \label{fig:n0}
\end{figure}

\begin{figure}
\centering
\includegraphics[clip=true,width=0.95\hsize]{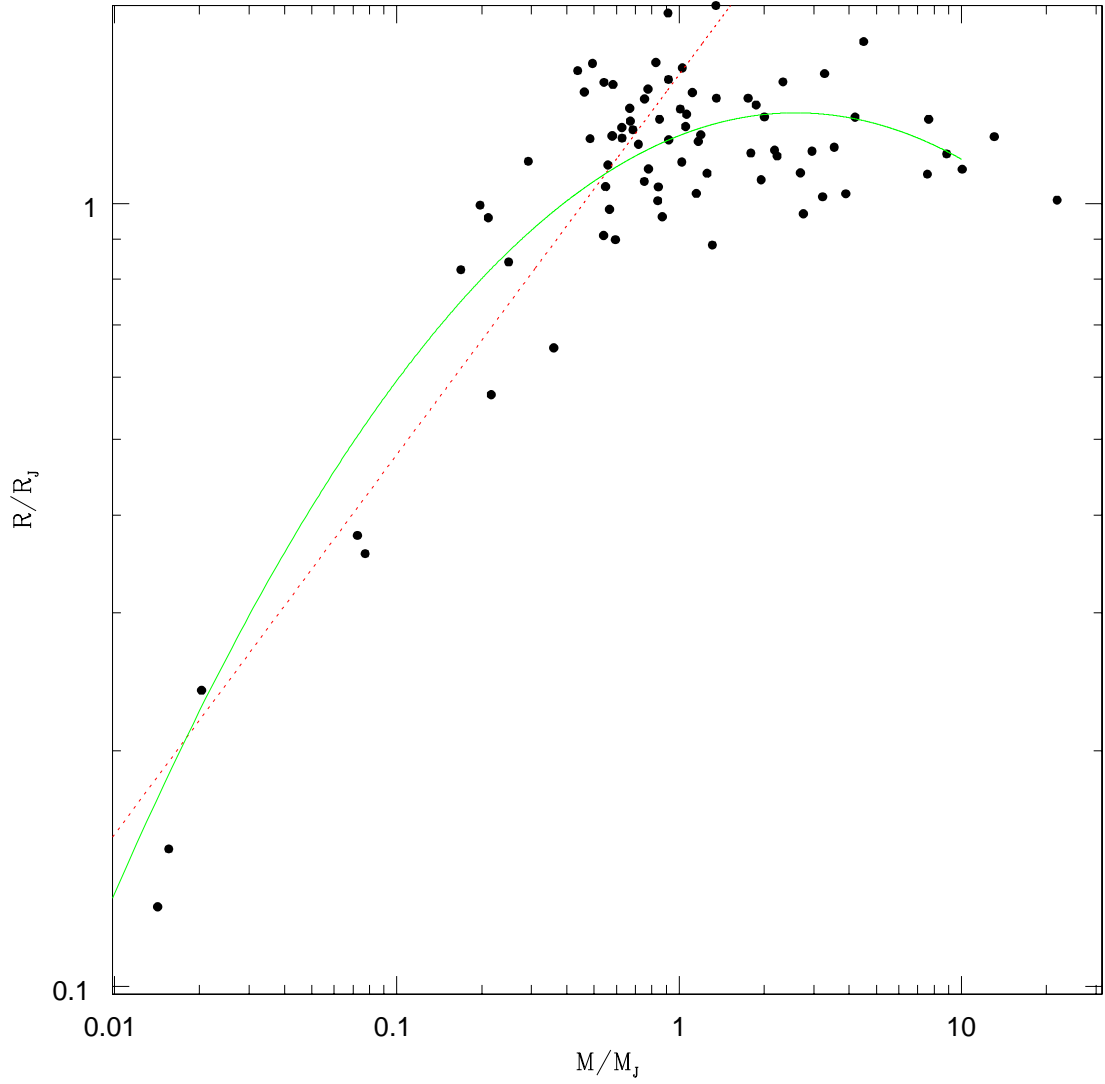}
\caption{The masses and radii of confirmed transiting exoplanets. The
  green solid line is the log-quadratic fit in equation (\ref{eq:mr}). The
  red dashed line is the log-linear fit
  $\log(M/M_\oplus)=2.06\log(R/R_\oplus)$ from \cite{liss11b}.}
\label{fig:mr}
\end{figure}

\noindent (ii) We may estimate $n_\mathrm{tot}^\mathrm{RV}$ using the
tranet frequency derived from the Kepler mission. Once again, we
restrict the Kepler sample to host stars that are F, G, and K dwarfs
(4000\,K$<T_\mathrm{eff} < $\,6500\,K and $\log g > 4$). We then carry
out the following steps for a given orbital period $P$ and velocity
semi-amplitude $K_\mathrm{RV}$: (i) compute the corresponding mass
$M(P,K_\mathrm{RV})=M_{J}(K_\mathrm{RV}/30\mbox{\,m
  s}^{-1})(1\mbox{\,yr}/P)^{1/3}$ assuming a circular orbit and a
solar-mass host star; (ii) find the number
$n^\mathrm{RV}(P,K_\mathrm{RV})$ of RV planets with period less than
$P$ and mass greater than $M(P,K_\mathrm{RV})$; (iii) find all Kepler
tranets with mass greater than $M(P,K_\mathrm{RV})$ and period less
than $P$, using an empirical mass-radius relation found by fitting
mass and radius measurements from transiting planets in the range
0.1--$10M_J$ (see Figure \ref{fig:mr}) to a log-quadratic relation
\begin{equation}
\log R/R_J=0.087
+0.141\log M/M_{J}-0.171\left(\log M/M_{J}\right)^2;
\label{eq:mr}
\end{equation} 
(iv) compute the total number of Kepler planets in this range
$n^\mathrm{Kep}(P,K_\mathrm{RV})$ by counting each tranet as $\epsilon^{-1}$
planets, to correct for geometric selection effects (eq.\
\ref{eq:goneone}); (v) estimate the total number of RV host stars as
$n_\mathrm{tot}^\mathrm{RV}=n_\mathrm{tot}^\mathrm{Kep}n^\mathrm{RV}(P,K)/n^\mathrm{Kep}(P,K)$.
The results are shown in Figure \ref{fig:n0} for $K_\mathrm{RV}=10, 15, 20,
25\,\mbox{m s}^{-1}$. As the majority of RV surveys have reached
precisions of $\sim 15\mbox{\,m s}^{-1}$ or better over the last
decade, it is reassuring but not surprising that the estimates of
$n_\mathrm{tot}^\mathrm{RV}$ for $K_\mathrm{RV}=15,20,25\mbox{\,m s}^{-1}$ are
consistent. The rise in $n^\mathrm{RV}_\mathrm{tot}$ at small periods
is likely due to the known discrepancy in hot Jupiter frequency
between transit and RV surveys (the frequency of hot Jupiters
estimated from transit surveys is factor of $\sim 2$ smaller than that
derived from RV surveys, perhaps because the average metallicities are
different; see \citealt{gould, howard}).

These independent approaches yield
$n^\mathrm{RV}_\mathrm{tot}\simeq2500\pm1000$ and
$n^\mathrm{RV}_\mathrm{tot}\simeq3000\pm1000$, respectively, which are
consistent within the errors. The corresponding inclination ranges
from Figure \ref{fig:rv_test} are $0<\langle\sin^2i\rangle^{1/2}<0.08$
and $0.02<\langle\sin^2i\rangle^{1/2}<0.09$ which correspond to an rms
or mean inclination range of 0--$5^\circ$ (as shown in
\S\ref{sec:fisher}, for a Rayleigh distribution the rms inclination is
only larger than the mean inclination by 12\%, which is much less than
the uncertainty).

The success of the separability assumption in modeling survey
selection effects (\S\ref{sec:survey} and Fig.\ \ref{fig:one}) suggests
that our results should be insensitive to cuts made on the Kepler
planet candidates. To check this, we have repeated the analysis for
the Kepler sample examined by \cite{liss11b}, who imposed a period cut $3\mbox{\,d}<P<125\mbox{\,d}$, a
radius cut $1.5R_\oplus\le R\le 6R_\oplus$, and a signal/noise cut
SNR$\,\ge16$, which reduced the number of planets to 63\% of our
sample. We find the mean inclination for this sample to be
0--$4^\circ$, not significantly different from the estimate 
in the preceding paragraph. 

Although the range of rms inclinations is tightly constrained by this
analysis, the multiplicity function is not. For example, within
1--$\sigma$ of the maximum-likelihood model ($\Delta\log P\le0.5$) we
have found models that have no 1-planet systems (67\% have no planets, 29\% have 2
planets, and 4\% have 13 planets) and others that have no zero-planet
systems (93\% have 1 planet, 2\% have 6 planets, and 5\% have 25
planets).

A by-product of this analysis is the ratio
$W^\mathrm{RV}/W^\mathrm{Kep}$ (eq.\ \ref{eq:cccca}), which measures
the relative sensitivity of the RV and Kepler surveys. This ratio varies
smoothly from 0.5 for razor-thin systems to 0.2 for
$\langle\sin^2i\rangle^{1/2}=0.1$, independent of the maximum number of
planets in the model. In other words 20--50\% of the Kepler planets could have
been detected in RV surveys. If this ratio can be determined
independently by fitting models of the period, radius, and mass
distributions it will provide a constraint on the rms inclination that
does not require estimating the total number of RV target stars. 

A weak link in these arguments is the assumption that the population of
FGK dwarf stars is
the same in the Kepler and RV surveys. One sign that these populations
are different is the higher frequency of hot Jupiters found in RV
surveys, as mentioned above. However, we note that our two approaches to estimating
$n^\mathrm{RV}_\mathrm{tot}$, one using only RV surveys and one
comparing the Kepler and RV surveys, yield similar answers, which
suggests that the estimate of the rms inclination that we derive from
this answer is insensitive to differences between the host stars of
the Kepler and RV surveys.
 
It is interesting to compare this estimate of the mean inclination to
the mean eccentricity for Kepler planets. Restricting our sample to
planets with minimum mass between 0.01 and 0.1 Jupiter masses and
period $P>10\mbox{\,d}$ (to avoid the effects of tidal
circularization), the mean eccentricity of planets discovered in RV
surveys is 0.15 (we have also excluded planets with a reported
eccentricity of zero, which may include cases in which no eccentricity
was fit). These results are roughly consistent with estimates of the
mean eccentricity of Kepler planets from transit timing: \cite{moor11}
find that the mean eccentricity is between 0.13 and 0.25 at a
$p$-value of 0.05. We have
\begin{equation}
\frac{\langle i\rangle}{\langle e\rangle}=0.35\frac{\langle
  i\rangle}{3^\circ}\frac{0.15}{\langle e\rangle}.
\end{equation}
Theoretical studies of eccentricity and inclination growth in
planetesimal disks \citep[e.g.,][]{ida} find $\langle i\rangle/\langle
e\rangle=0.45$--0.5, somewhat larger than this value. A possible
explanation is that the eccentricities may have been systematically
overestimated. \cite{zak} find that the typical bias due to
measurement errors is $\Delta e\sim 0.04$ in RV catalogs, and the bias
in this sample is likely to be higher since the SNR is low for
low-mass planets. Possibly a similar bias is present in the Kepler
measurements of the eccentricity distribution.

The Kepler survey can measure transit timing variations of a minute or
less in favorable cases \citep{ford11}. These variations can be used
to detect and characterize additional planets.  Given the rms
inclination of 0--0.09 radians that we have derived, roughly 20--30\%
of the single-tranet Kepler systems are expected to have additional
planets (Figure \ref{fig:total}), and many of these may be detectable
by transit timing variations. \cite{ford11} estimate that
$\sim10$--$20\%$ of suitable Kepler tranets show evidence 
of transit timing variations, and this number is likely to increase as
the survey duration grows.  Figure \ref{fig:total} also shows that the
fraction of two- or three-tranet systems with additional planets is
substantially higher, and strongly dependent on the rms inclination. A
preliminary analysis by \cite{ford11} yields much lower probabilities
of 0.1--0.2 for two- and three-tranet systems; such low probabilities
would be difficult to reconcile with any of our models, whatever the rms
inclination may be.

\section{Summary}

\noindent
We have described a methodology for analyzing the multiplicity
function---the fraction of host stars containing a given number of
planets---in radial-velocity (RV) and transit surveys. Our approach is
based on the approximation of separability, that the probability
distribution of planetary parameters in an $n$-planet system is the
product of identical 1-planet distributions
(\S\ref{sec:separable}). Exoplanet surveys show that separability is
not precisely satisfied but the departures from this approximation are
small enough that it provides a powerful tool for the study of
multi-planet systems. Using this approximation we have shown how to
relate the observable multiplicity function in surveys with different
sensitivities, so long as they examine populations of potential host
stars with similar properties (\S\ref{sec:survey}).  We have also
shown how to derive the multiplicity function from transit surveys
(\S\ref{sec:transit}) assuming a given form for the inclination
distribution (the Fisher distribution, \S\ref{sec:fisher}). Our
principal conclusions are:

\begin{enumerate}

\item At present, the Kepler data alone \citep{bor11} are not able to
  constrain the inclination distribution of multi-planet systems
  without additional assumptions or data. In particular, models with
  all rms inclinations---from razor-thin to spherical---are able to
  reproduce the observable multiplicity function in the Kepler sample. This
  conclusion differs from \cite{liss11b}, who found that (i) the Kepler
  data contained an excess of single-tranet systems that could not be
  fit by any of their models; (ii) models with mean inclinations
  exceeding $5^\circ$ were poor fits to the data. We believe that
  these conclusions reflect the restricted, though plausible, range of
  models for the multiplicity function considered by \cite{liss11b},
  although their estimated upper limit to the mean inclination is
  entirely consistent with our conclusions below based on other methods.

\item Systems with large rms inclinations are only consistent with the
  Kepler data if at least some of them contain a large number of planets. The relation
  between rms inclination and maximum number of planets is given by
  equation (\ref{eq:maxp}). 

\item In our models, the percentage of one-tranet systems with
  additional planets is 20--30\%, and for two- or three-tranet systems
  this percentage is even higher (Figure \ref{fig:total}). These
  fractions can be probed observationally using transit timing
  variations. 

\item The rms inclination can be constrained by combining estimates
  of the observable multiplicity function from Kepler and RV surveys, but only after
  estimating the effective number of stars that have been examined in
  RV surveys. We have made two estimates, one using Kepler data and
  one without; these are consistent, and yield
  $\langle\sin^2i\rangle^{1/2}\le 0.09$,
  corresponding to mean inclinations in the range 0--5$^\circ$. 

\item Although the range of rms inclinations is tightly constrained by
  this analysis, the multiplicity function
  is not: the data are well-fit by (presumably) pathological models
  containing no zero-planet systems, no one-planet
  systems, etc. 

\end{enumerate}

This research was supported in part by NASA grant NNX08AH83G, and has
made use of the Exoplanet Orbit Database and the Exoplanet Data
Explorer at exoplanets.org. Work by SD was performed under contract
with the California Institute of Technology (Caltech) funded by NASA
through the Sagan Fellowship Program. We acknowledge helpful
conversations with Dan Fabrycky, Debra Fischer, Matt Holman, Boaz
Katz, Darin Ragozzine, and Jason Wright.

\end{document}